\documentclass[twocolumn,prb,superscriptaddress]{revtex4-1}
\usepackage[utf8x]{inputenc}
\usepackage[english]{babel}
\usepackage{mathtools}
\usepackage{amsmath}
\usepackage{amssymb}
\usepackage{graphicx}
\usepackage{hyperref}
\usepackage{txfonts}
\usepackage[usenames,dvipsnames]{color}

\begin{document}

\title{Role of coherence in transport through engineered atomic spin devices}
\author{Alexey M. Shakirov}
\email[]{a.shakirov@rqc.ru}
\affiliation{Russian Quantum Center, Novaya street 100A, 143025 Skolkovo, Moscow
Region, Russia}
\affiliation{Department of Physics, Lomonosov Moscow State University, Leninskie
gory 1, 119992 Moscow, Russia}
\author{Yulia E. Shchadilova}
\affiliation{Russian Quantum Center, Novaya street 100A, 143025 Skolkovo, Moscow
Region, Russia}
\affiliation{Department of Physics, Harvard University, Cambridge, Massachusetts
02138, USA}
\author{Alexey N. Rubtsov}
\affiliation{Russian Quantum Center, Novaya street 100A, 143025 Skolkovo, Moscow
Region, Russia}
\affiliation{Department of Physics, Lomonosov Moscow State University, Leninskie
gory 1, 119992 Moscow, Russia}
\author{Pedro Ribeiro}
\affiliation{CeFEMA, Instituto Superior Técnico, Universidade de Lisboa, Av. Rovisco
Pais, 1049-001 Lisboa, Portugal}
\affiliation{Russian Quantum Center, Novaya street 100A, 143025 Skolkovo, Moscow
Region, Russia}

\begin{abstract}
We give a further step in the quantum mechanical description of engineered atomic spin structures by deriving a master equation of the Redfield type that governs the dynamics of the atomic spin density matrix.
By generalizing this approach to charge-specific density matrices, we are able to describe magnetic transport quantities, such as the average inelastic current and the shot noise, accessible by tunneling spectroscopy.
Our method suitably describes moderate lead-atom coupling regimes where quantum coherence effects cannot be disregarded.
We contrast our approach with the existing descriptions in terms of rate equations and show examples where coherence effects are crucial to understand the physics of spin-polarized tunnel current through spin structures.
\end{abstract}

\maketitle

\global\long\def\ket#1{\left| #1\right\rangle }

\global\long\def\bra#1{\left\langle #1 \right|}

\global\long\def\kket#1{\left\Vert #1\right\rangle }

\global\long\def\bbra#1{\left\langle #1\right\Vert }

\global\long\def\braket#1#2{\left\langle #1\right. \left| #2 \right\rangle }

\global\long\def\bbrakket#1#2{\left\langle #1\right. \left\Vert #2\right\rangle }

\global\long\def\av#1{\left\langle #1 \right\rangle }

\global\long\def\tr{\text{Tr}}

\global\long\def\pd{\partial}

\global\long\def\im{\text{Im}}

\global\long\def\re{\text{Re}}

\global\long\def\sgn{\text{sgn}}

\global\long\def\Det{\text{Det}}

\global\long\def\abs#1{\left|#1\right|}

\global\long\def\up{\uparrow}

\global\long\def\down{\downarrow}

\global\long\def\bs#1{\boldsymbol{#1}}

\section{Introduction}
\label{sec:Introduction}

Advances in the field of low-temperature scanning tunneling microscopy (STM) have enabled the detection and manipulation of the spin of individual magnetic atoms and molecules\cite{Stipe1998}.
With current STM techniques magnetic atoms can be arranged into artificial assemblies such as chains, ladders or few-atom aggregates \cite{Hirjibehedin2006,Loth2010,Gauyacq2012,Spinelli2014}, hereafter referred to as engineered atomic spin devices (EASDs).
The ability to manipulate and monitor individual atomic spins using inelastic electron tunneling spectroscopy has permitted us to address a set of new questions such as the origin and nature of magnetism in few-atom aggregates and nanostructures, the effects of many-particle correlations between the localized atomic spins and the itinerant electrons crossing the system, and the identification of spin excitations from differential conductance spectra.
In parallel to fundamental physics aspects, EASDs are of major interest for spintronic applications \cite{Leuenberger2001,Troiani2005,Imre2006,Khajetoorians2011}.

Up to now, EASDs have mostly been applied to improve classical information storage technology.
However, as the exploration of coherent quantum regimes is becoming experimentally reachable, these devices are of great potential for applications in quantum information processing and manipulation.

A typical EASD consists of a set of magnetic atoms deposited on a crystalline few-atoms-thick layer of insulating material that coats a metallic substrate (see Fig. \ref{fgr:Sketch}).
The presence of the insulator reduces the hybridization of the atoms with the underlying metallic substrate and strongly suppresses charge fluctuations.
This leaves the atomic spin as the only relevant low-energy degree of freedom. 
Each atom can be addressed individually by a metallic spin-polarized STM tip. 
An electronic current ensues by applying a finite bias voltage between the substrate and the tip, collecting contributions from elastic and inelastic processes.
Elastic processes arise when electrons pass from one metallic lead to the tip with no energy change.
They can be due to direct tip-substrate hopping, amounting to a trivial contribution to the differential conductance, or due to mediated hopping via degenerate energy states of the atomic structure -- the mechanism responsible for Kondo-like physics \cite{Hewson1997}. However, for temperatures or voltages larger than the Kondo energy scale, nontrivial elastic processes can be neglected.
Inelastic processes arise when the electrons, while tunneling through the atomic structure, exchange energy with its internal degrees of freedom. 

The theory of inelastic tunneling through EASDs has received important contributions in recent years.
A perturbative approach, assuming small tip-atom and substrate-atom couplings, was developed \cite{Fransson2009,Fernandez-Rossier2009,Delgado2010a,Ternes2015} in parallel with a strong-coupling approach \cite{Persson2009,Lorente2009}.
These approaches, based on a set of classical rate equations, predict the current-voltage characteristics of the system.
In particular, they model the signature of the atomic structure excitation spectrum in the measured differential conductance \cite{Otte2009,Delgado2010a}.
Despite these substantial advances, a complete picture of the nonequilibrium transport processes in EASDs is still far from complete.
A particular aspect that requires better understanding is the role of nondiagonal components of the density matrix, i.e., quantum coherences.
Existing works mostly concentrate on the computation of the decoherence times \cite{Delgado2010,Gauyacq2015,Delgado2016}, leaving out the question of the effect of coherence in the observables.
This issue is of major importance if EASDs are to be operated in quantum coherent regimes, e.g., as devices for quantum information processing.

In this work we give the first steps in the direction of a quantum mechanical description of the dynamics in EASDs.
We use a theoretical approach based on the microscopically derived Redfield equation \cite{Pollard1996,Breuer2002} for the density matrix of the atomic subsystem. 
The Redfield equation is a type of master equation describing the evolution of an open quantum system weakly coupled to its environment.
Originally employed to model nuclear magnetic resonance \cite{Wangsness1953,Bloch1957,Redfield1957}, it has been applied in various fields including quantum optics \cite{Scully1997,Breuer2002,Gardiner2004}, chemical dynamics \cite{Nitzan2006}, and electronic transport \cite{Esposito2009}.

Our goal is to describe inelastic transport processes in EASDs, in particular to predict the average value of the current and the shot noise measured by STM. 
To access the information about the electronic current through the system, we generalize the Redfield equation approach to charge-specific density matrices \cite{Rammer2004,Flindt2004,Flindt2005}.
We derive expressions for the steady state values of the average current and for the shot noise.
In order to illustrate our method, we consider single atoms of different total spin and an atomic chain as examples.
We study how coherences affect the current and shot-noise characteristics for several setups including different tip polarization geometries.
The results are compared with the previous approaches where coherences are neglected \cite{Fernandez-Rossier2009,Delgado2010a} in order to highlight regimes where coherent dynamics sets in. 

The paper is organized as follows.
Section \ref{sec:Model} gives a description of the setup and the model Hamiltonian. 
Section \ref{sec:Method} describes the method.
Section \ref{sec:MethodSummary} summarizes the methodology and provides the final expressions for the average current and the shot noise.
The details of the derivation are presented in Sec. \ref{sec:MethodDetails} and the application to EASDs is given in Sec. \ref{sec:Application}.
In Sec. \ref{sec:Results} we present some illustrative examples: a single atom with spin $1/2$, a single atom with spin $5/2$, and a chain of atoms with spin $1/2$.
We discuss our results and draw conclusions in Sec. \ref{sec:Discussion}. 
Appendices are devoted to technical details of the derivation.

\section{Model}
\label{sec:Model}

\begin{figure}[t]
\includegraphics[width=0.8\columnwidth]{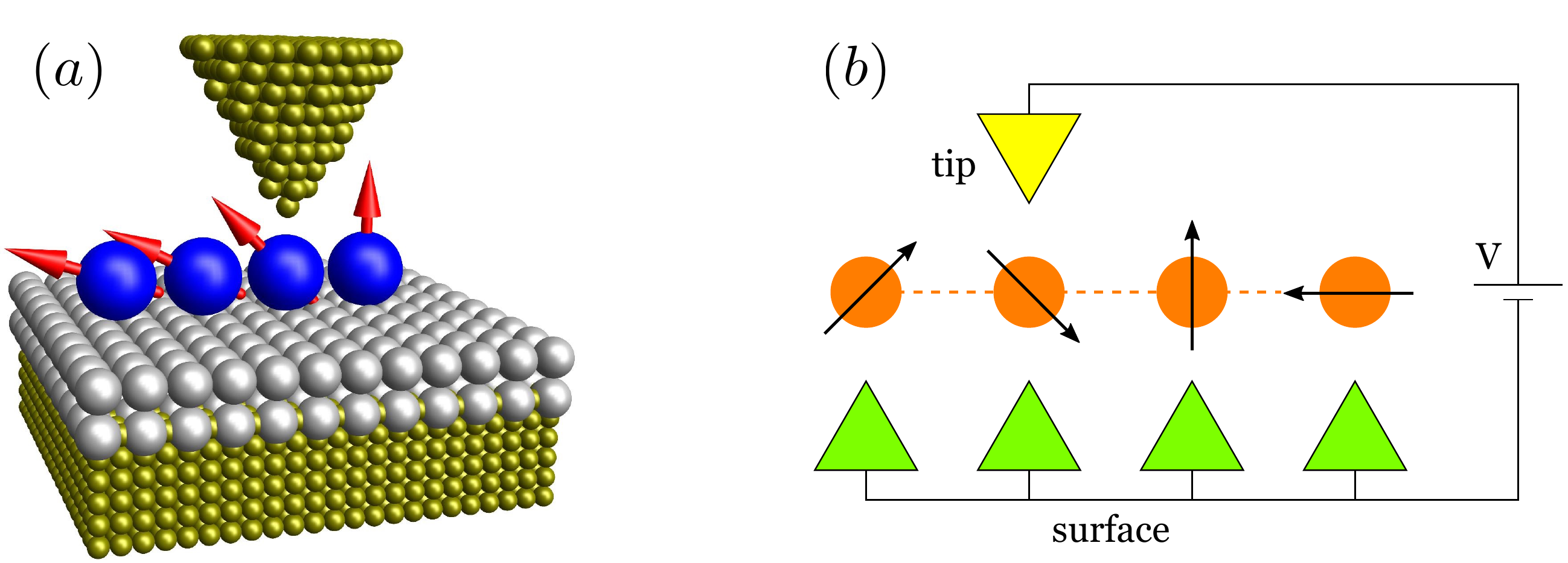}
\caption{(a) Schematics of a typical EASD.
Magnetic atoms are deposited on an insulating layer coating a metallic substrate.
Upon applying a voltage difference between the metallic STM tip and the substrate, a charge current ensues.
(b) Sketch of the model.
The substrate is modeled by a set of independent reservoirs sharing the same chemical potential $\mu_{S}$.
The tip is modeled by an additional reservoir with $\mu_{T}=\mu_{S}-eV$.
All reservoirs are assumed to be wideband metals.}
\label{fgr:Sketch} 
\end{figure}

A generic setup of an EASD, sketched in Fig. \ref{fgr:Sketch}, can be described by the Hamiltonian $ {H}= {H}_{A}+H_{R}+ {H}_{I}$, which includes the Hamiltonian of the atomic subsystem $H_{A}$, the Hamiltonian of the electronic degrees of freedom of the tip and of the substrate $H_{R}$, and the coupling Hamiltonian $H_{I}$.
In the following we specify and describe each term. 

\paragraph*{Magnetic atoms.}

We consider the limit when the atomic charge gap is much larger than other characteristic energies. 
The atoms thus possess a well defined number of electrons, and tunneling through atomic orbitals is only possible by virtual excitations of different charge states. 
Therefore, each atom behaves as a localized spin coupled to other atoms and to the spin of conduction electrons by an effective exchange term \cite{Anderson1966}.
As a result, the low-energy Hamiltonian of the atomic ensemble can be expressed solely in terms of spin degrees of freedom \cite{Gatteschi2006}, with symmetry arguments dictating its generic form \cite{Hirjibehedin2007,Otte2008,Fernandez-Rossier2009}
\begin{equation}
\begin{split}
&H_{A}=\sum_{r}\left[DS^{2}_{rz'}+E\left(S^{2}_{rx'}-S^{2}_{ry'}\right)\right]+\\
&+\sum_{\langle rr'\rangle}J_{rr'}\mathbf{S}_{r}\cdot\mathbf{S}_{r'}+g\mu_{B}\sum_{r}\mathbf{B}\cdot\mathbf{S}_{r},
\end{split}
\label{eqn:Cluster}
\end{equation}
where $r=1,...,L$ enumerates the atoms.
The first term describes the magnetic anisotropy of the crystal parametrized by the coefficients $D$ and $E$.
Here the spin is quantized along the principal axes of the crystal $x'$ (hard axis), $y'$ (intermediate axis), and $z'$ (easy axis).
The second term corresponds to an effective exchange $J_{rr'}$ between pairs of neighboring atoms $\langle rr'\rangle$ arising from, e.g., the superexchange, or the RKKY interaction mediated by the substrate.
The third term is the Zeeman splitting induced by an external magnetic field $\mathbf{B}$ and proportional to the atomic $g$ factor.

\paragraph*{Substrate and tip.}

We model the substrate as a set of identical metallic reservoirs each one coupled to a single atom (see Fig. \ref{fgr:Sketch}).
This describes the limit when the substrate-mediated correlations between the atoms, other than the effective exchange interaction, are negligible.
The polarized tip is modeled as an additional metallic reservoir coupled to a specific atom $r_{0}$.
For an ensemble of $L$ atoms, this amounts to considering an environment consisting of $L+1$ electronic reservoirs in total. 
The Hamiltonian of the reservoirs is given by $H_{R}=H_{T}+ H_{S}$ with the corresponding tip and substrate Hamiltonians
\begin{equation}
H_{T}=\sum_{\sigma k}\varepsilon_{\sigma k}^{(T)}f_{\sigma k}^{\dag}f_{\sigma k},~~~H_{S}=\sum_{r\sigma k}\varepsilon_{\sigma k}^{(S)}c_{r\sigma k}^{\dag}c_{r\sigma k},
\end{equation}
where $\sigma=\uparrow,\downarrow$ is the spin of electrons quantized along the tip polarization vector $\mathbf{P}$, and $k$ runs over single-particle states of the reservoirs.
Electrons in all reservoirs (tip and substrate) are in thermal equilibrium with a common temperature $1/\beta$ (in energy units) and chemical potentials $\mu_{S}$ for the substrate and $\mu_{T}=\mu_{S}-eV$ for the tip, where $V$ is the applied voltage and $-e$ is the electron charge.
The metallic nature of the electronic reservoirs translates to a local density of states $\varrho_{\eta\sigma}(\varepsilon)=\mathcal{V}_{\eta}^{-1}\sum_{k}\delta\left(\varepsilon-\varepsilon_{\sigma k}^{(\eta)}\right)$ with $\eta=T,S$, that may be considered energy-independent within the energy scales of interest.
Here $\mathcal{V}_{\eta}$ stands for the volume of the reservoir.
We introduce the spin-dependent density of states to account for the tip polarization.
For electrons in the tip we assign $\varrho_{T\sigma}=w_{\sigma}\varrho_{T}$, with $w_{\uparrow}=1+p$, $w_{\downarrow}=1-p$, where $p$ is the polarization parameter ranging from $-1$ to $1$.
For electrons in the unpolarized substrate $\varrho_{S\uparrow}=\varrho_{S\downarrow}=\varrho_{S}$.
Even though we work in the wideband approximation, for regularization purposes we use rectangular-shaped densities of states
\begin{equation}
\varrho_{\eta}(\varepsilon)=\varrho_{\eta}\Theta\left(W-|\varepsilon|\right),
\end{equation}
where $\Theta$ is the Heaviside function and $W$ is the bandwidth, much larger than other energy scales of the system.

\paragraph*{Coupling.}
The coupling of the atoms to the electrons in the leads is described by the exchange interaction Hamiltonian \cite{Appelbaum1967,Kim2004,Fernandez-Rossier2009,Delgado2010} $H_{I}=\sum_{\eta\eta'}H_{\eta\eta'}$ with
\begin{equation}
\begin{split}
&  H_{TS}=\sqrt{J_{T}J_{S}}\sum_{a\sigma\sigma'kk'} {S}_{r_{0}a}\otimes c_{r_{0}\sigma k}^{\dag}\tau_{\sigma\sigma'}^{a}f_{\sigma'k'},\\
&  H_{TT}=J_{T}\sum_{a\sigma\sigma'kk'} {S}_{r_{0}a}\otimes f_{\sigma k}^{\dag}\tau_{\sigma\sigma'}^{a}f_{\sigma'k'},\\
&  H_{SS}=J_{S}\sum_{ra\sigma\sigma'kk'} {S}_{ra}\otimes c_{r\sigma k}^{\dag}\tau_{\sigma\sigma'}^{a}c_{r\sigma'k'},
\end{split}
\label{eqn:Coupling}
\end{equation}
where $J_{\eta}\simeq2u_{\eta}^{2}U\Delta_{\eta}^{-1}\left(\Delta_{\eta}+U\right)^{-1}$ are the exchange coupling energies determined by the lead-atom hopping amplitude $u_{\eta}$, the intra-atomic Coulomb repulsion $U$ between electrons, and the energy difference $\Delta_{\eta}$ between the atomic level and the Fermi energy of the lead \cite{Schrieffer1966}.
$\tau^{a}$ and $ {S}_{ra}$ with $a=x,y,z$, are the Pauli matrices and the spin operators of the atom $r$, respectively.
The axes are chosen so that $z$ is aligned with the tip polarization $\mathbf{P}$.
The inelastic current through the cluster originates from tip-to-substrate $H_{TS}$ and substrate-to-tip $H_{ST}=H_{TS}^{\dag}$ tunneling, while the terms $H_{TT}$ and $H_{SS}$ yield purely relaxational contributions due to tip-to-tip and substrate-to-substrate electron scattering processes.
In Eq. (\ref{eqn:Coupling}) we have neglected momentum dependence of the lead-atom hopping amplitude and used spin rotational invariant exchange coupling.
In the following we use dimensionless parameters $\gamma_{\eta}=\pi J_{\eta}\varrho_{\eta}\mathcal{V}_{\eta}$ to characterize the strength of the tip-atom and substrate-atom couplings.

\section{Method}
\label{sec:Method}

\subsection{Summary}
\label{sec:MethodSummary}

In this section we summarize the main results of our approach to the description of the transport and dynamics in EASD setups.
We discuss the properties of the master equation governing the dynamics of the atomic subsystem and present the generic expressions for the average value of the inelastic current and the shot noise.

\paragraph*{Master equation.}

Following a standard set of approximations \cite{Breuer2002,Rammer2004} (see below), we derive a Redfield-type master equation for the  density matrix of the atomic subsystem 
\begin{equation}
\partial_{t} {\rho}=\mathcal{L} {\rho},
\label{eqn:summaryME}
\end{equation}
where the superoperator $\mathcal{L}$ is given in Eq. (\ref{eqn:LsuperEASD}).
The steady state density matrix $ {\rho}_{\infty}$ is calculated as the eigenstate of $\mathcal{L}$ corresponding to zero eigenvalue, i.e., $\mathcal{L} {\rho}_{\infty}=0$.

The derivation of Eq. (\ref{eqn:summaryME}) assumes the lead-atom coupling to be small within the Born approximation and the leads to have a short memory time. 
Nonetheless, although a Markov-like  approximation is employed, the Redfield equation does not lead to purely Markovian evolution \cite{Wolf2008,Breuer2009,Rivas2010,Hall2014,Ribeiro2015}. 
Therefore, the Redfield equation is generally not of the Lindblad form and may violate the positivity of the density matrix \cite{Alicki2007}.
To prevent the breakdown of positivity, the rotating wave approximation (RWA) is sometimes performed leading to an equation where the dynamics of populations and coherences decouple \cite{Breuer2002}, which implies that the coherences vanish in the steady state.
This further approximation is valid when the damping rate is much slower than the Bohr frequencies of the system and is equivalent to a treatment in terms of rate equations. 
Neglecting coherences may lead to wrong predictions when they become of the same order as populations \cite{Kaiser2006,Harbola2006}.
On the other hand, the violation of positivity during the dynamics by the Redfield equation generally occurs only far from equilibrium; the description of the stationary regime is in general accurate given that the density matrix remains physical  \cite{Pechukas1994}.

In the present case, at low temperatures as compared with the energy scales of the atomic spin system, this approach is valid as long as the lead-atom coupling is moderate.
Away from its range of validity, the steady-state density matrix of Eq. (\ref{eqn:summaryME}) may violate positivity yielding to unphysical results. 
For vanishing coupling we recover the rate equation results for the evolution of the populations.
The method thus suitably describes moderate lead-atom coupling regimes where coherences cannot be disregarded.
In our numerical studies below we explicitly checked that $\rho_{\infty}$ is a physically sound density matrix, i.e., has no negative eigenvalues.

The approach followed here, due to its perturbative nature, is unable to capture nonperturbative phenomena in the lead-atom coupling, e.g., elastic processes responsible for the Kondo-like physics when the atomic structure has a degenerate ground-state manifold.
Here we assume nontrivial elastic processes to be absent either by considering nondegenerate atomic spectra or by assuming temperature regimes where such effects are washed away. 

\paragraph*{Current and shot noise.}

In order to describe transport properties, Eq. (\ref{eqn:summaryME}) has been generalized to describe the evolution of charge-specific density matrices (CSDMs) that describe the state of the system given that a certain number of charge carriers has left the tip.
Using the method of CSDMs \cite{Rammer2004,Flindt2004,Flindt2005}, we obtained the expression for the average value of the inelastic current in the steady state as
\begin{equation}
I=-e\,\mbox{tr}\left(\mathcal{J} {\rho}_{\infty}\right),
\end{equation}
where the current superoperator $\mathcal{J}$ is defined in Eqs. (\ref{eqn:DJ-definition}) and (\ref{eqn:DsuperEASD}).
Elastic terms, appearing in the current spectra due to direct tunneling of electrons between the tip and the substrate, are not accounted in this expression.
These contributions have no impact on the dynamics of the atoms and can be calculated independently.  

The shot noise of the inelastic current in the steady state can be expressed as \cite{Flindt2004,Flindt2005}
\begin{equation}
S=4e^{2}\mbox{tr}\left(\mathcal{D} {\rho}_{\infty}-\mathcal{J}\mathcal{L}^{-1}\mathcal{J} {\rho}_{\infty}\right),
\label{eqn:summaryShotNoise}
\end{equation}
where $\mathcal{L}^{-1}$ is the pseudoinverse of $\mathcal{L}$, and the superoperator $\mathcal{D}$ is defined in Eqs. (\ref{eqn:DJ-definition}) and (\ref{eqn:DsuperEASD}).

The above set of expressions allows us to reproduce the results of Sec. \ref{sec:Results} and is given here for the benefit of a reader who might not be interested in the detailed derivation of the method. 

\subsection{Derivation}
\label{sec:MethodDetails}

In this section we provide a derivation of the master equation for a generic system, as well as expressions for the current and the shot noise in the steady state.
Our approach is based on the master equations for CSDMs introduced in Ref. \cite{Rammer2004} for an open quantum system driven by a particle flow.
In Ref. \cite{Rammer2004} the authors consider a system coupled to two reservoirs (here identified as tip and substrate) with a coupling Hamiltonian $H_{I}$ containing the terms $H_{TS}$ and $H_{ST}$ of Eq. (\ref{eqn:Coupling}).
Here we generalize this approach to include relaxation processes due to tip-to-tip and substrate-to-substrate scattering of the electrons, i.e., terms $H_{TT}$ and $H_{SS}$ in Eq. (\ref{eqn:Coupling}).
Not to restrict the derivation to our particular spin system, in this section we write 
\begin{equation}
H_{\eta\eta'}=\sqrt{J_{\eta}J_{\eta'}}\sum_{\alpha\alpha'} {T}_{\alpha\alpha'}c_{\alpha}^{\dag}c_{\alpha'},
\label{eqn:R-operators}
\end{equation}
where $ {T}_{\alpha\alpha'}$ are generic operators of the atomic subsystem, and index $\alpha$ parametrizes quantum numbers of the electrons in the substrate ($\eta=S$) or the tip ($\eta=T$), i.e., $\alpha=(\sigma,k)$ for $\eta=T$ and $\alpha=(r,\sigma,k)$ for $\eta=S$.
The identification of $ {T}_{\alpha\alpha'}$ with specific spin operators of the magnetic atoms is done in Sec. \ref{sec:Application}.

\subsubsection{Charge-specific density matrices}

CSDMs $ {\rho}_{n}$ of the atomic subsystem are defined as 
\begin{equation}
 {\rho}_{n}=\mbox{tr}_{R}\left(\mathcal{P}_{n}\rho_{\text{tot}}\right),
\label{eqn:CSDMdef}
\end{equation}
where $\rho_{\text{tot}}$ is the full density matrix of the system (atoms plus leads) and $\mbox{tr}_{R}$ stands for the trace over all reservoirs.
The operator $\mathcal{P}_{n}$ projects the full Hilbert space into a subspace with $n$ particles transferred from the tip to the substrate (compared to the initial state).
Note that summing up CSDMs recovers the density matrix of the system $\sum {\rho}_{n}= {\rho}$.
As shown in Appendix \ref{sec:Derivation}, they evolve according to the equations of motion 
\begin{equation}
\partial_{t} {\rho}_{n}+i\left[H_{A}, {\rho}_{n}\right]=-i\mbox{tr}_{R}\left(\mathcal{P}_{n}\left[H_{I},\rho_{\text{tot}}\right]\right).
\label{eqn:EOMinitial}
\end{equation}
The substitution of $H_{I}=\sum_{\eta\eta'}H_{\eta\eta'}$ into the right-hand side of Eq. (\ref{eqn:EOMinitial}) leads to
\begin{equation}
\begin{split}
&\partial_{t} {\rho}_{n}+i\left[ {H}_{A}, {\rho}_{n}\right]=-i\left[\sum_{\eta\alpha}J_{\eta}f_{\alpha} {T}_{\alpha\alpha}, {\rho}_{n}\right]+\\
&+\sum_{\eta\eta'\alpha\alpha'}\sqrt{J_{\eta}J_{\eta'}}\left( {T}_{\alpha\alpha'} {C}_{\alpha\alpha'}^{(n)}+\mbox{h.c.}\right)
\end{split}
\label{eqn:EOMnonliouv}
\end{equation}
(see Appendix \ref{sec:Derivation}) with operators $ {C}_{\alpha\alpha'}^{(n)}$ defined as
\begin{equation}
i {C}_{\alpha\alpha'}^{(n)}=\mbox{tr}_{R}\left(\left(c_{\alpha}^{\dag}c_{\alpha'}-f_{\alpha}\delta_{\alpha\alpha'}\right)\rho_{\text{tot}}\mathcal{P}_{n}\right).
\label{eqn:C-operators}
\end{equation}
The numbers $f_{\alpha}=\langle c_{\alpha}^{\dag}c_{\alpha}\rangle$ are determined from the distribution function of electrons in the leads.
As shown in Appendix \ref{sec:C-operators}, the operators $ {C}_{\alpha\alpha'}^{(n)}$ satisfy the equations of motion 
\begin{equation}
\begin{split}
&\partial_{t} {C}_{\alpha\alpha'}^{(n)}+i[ {H}_{A}, {C}_{\alpha\alpha'}^{(n)}]-i\left(\varepsilon_{\alpha}-\varepsilon_{\alpha'}\right) {C}_{\alpha\alpha'}^{(n)}=\\
&=-\mbox{tr}_{R}\left(\left(c_{\alpha}^{\dag}c_{\alpha'}-f_{\alpha}\delta_{\alpha\alpha'}\right)\left[ {H}_{I},\rho_{\text{tot}}\right]\mathcal{P}_{n}\right).
\end{split}
\label{eqn:ExactEOMforC}
\end{equation}

\subsubsection{Approximations}

Up to this point all the equations were exact.
To proceed and obtain a closed set of equations for the evolution of CSDMs, a number of physically motivated approximations has to be made.
Following the standard derivation of the Redfield master equation \cite{Pollard1996,Breuer2002}, we employ both Born and Markov approximations.
Within these approximations components of the full density matrix $\mathcal{P}_{m}\rho_{\text{tot}}\mathcal{P}_{n}$ with $m\neq n$ vanish.
This is due to the fact that tunneling is rare and superpositions of states with different numbers of particles in the leads do not occur at this order in the lead-atom coupling.
For diagonal components we assume separability $\mathcal{P}_{n}\rho_{\text{tot}}\mathcal{P}_{n}\approx {\rho}_{n}\otimes\rho_{R}$ within the Born approximation.
This yields an approximate equation of motion for $C_{\alpha\alpha'}^{(n)}$ 
\begin{equation}
\begin{split}
&\partial_{t} {C}_{\alpha\alpha'}^{(n)}+i\left[ {H}_{A}, {C}_{\alpha\alpha'}^{(n)}\right]-i\left(\varepsilon_{\alpha}-\varepsilon_{\alpha'}\right) {C}_{\alpha\alpha'}^{(n)}\approx\sqrt{J_{\eta}J_{\eta'}}\times\\
&\times\left(\left(1-f_{\alpha}\right)f_{\alpha'} {\rho}_{n-n_{\alpha\alpha'}} {T}_{\alpha\alpha'}^{\dag}-f_{\alpha}\left(1-f_{\alpha'}\right) {T}_{\alpha\alpha'}^{\dag} {\rho}_{n}\right),
\end{split}
\label{eqn:ApproximateEOMforC}
\end{equation}
whose solution is given by 
\begin{equation}
\begin{split}
&C_{\alpha\alpha'}^{(n)}(t)=\sqrt{J_{\eta}J_{\eta'}}\int\limits_{0}^{t}e^{-i {H}_{A}\tau}\left(\left(1-f_{\alpha}\right)f_{\alpha'} {\rho}_{n-n_{\alpha\alpha'}}(t-\tau)\times\right.\\
&\left.\times {T}^{\dag}_{\alpha\alpha'}-f_{\alpha}\left(1-f_{\alpha'}\right) {T}^{\dag}_{\alpha\alpha'} {\rho}_{n}(t-\tau)\right)e^{i {H}_{A}\tau}e^{i(\varepsilon_{\alpha}-\varepsilon_{\alpha'})\tau}d\tau
\end{split}
\label{eqn:ApproximateEOMforCSolution}
\end{equation}
(see Appendix \ref{sec:C-operators}).
We assume that the memory time of the leads is short enough to extend the integration limit in the former expression to infinity.
Additionally, within the Born approximation we obtain 
\begin{equation}
e^{-i {H}_{A}\tau} {\rho}_{n}(t-\tau)e^{i {H}_{A}\tau}\approx {\rho}_{n}(t).
\label{eqn:ZeroApproxForCSDM}
\end{equation}
Then $ {C}_{\alpha\alpha'}^{(n)}$ are time-independent and expressed as 
\begin{equation}
\begin{split}
& {C}_{\alpha\alpha'}^{(n)}=\sqrt{J_{\eta}J_{\eta'}}\left(\left(1-f_{\alpha}\right)f_{\alpha'} {\rho}_{n-n_{\alpha\alpha'}} {\mathcal{T}}_{\alpha\alpha'}^{\dag}-\right.\\
&\left.-f_{\alpha}\left(1-f_{\alpha'}\right) {\mathcal{T}}_{\alpha\alpha'}^{\dag} {\rho}_{n}\right),
\end{split}
\label{eqn:Cfinal}
\end{equation}
where we have introduced the operators 
\begin{equation}
 {\mathcal{T}}_{\alpha\alpha'}=\int\limits _{0}^{\infty}e^{-i {H}_{A}\tau} {T}_{\alpha\alpha'}e^{i {H}_{A}\tau}e^{-i(\varepsilon_{\alpha}-\varepsilon_{\alpha'})\tau}d\tau.
\label{eqn:NewOps}
\end{equation}
In the eigenbasis $|m\rangle$ of $ {H}_{A}$, i.e., $ {H}_{A}|m\rangle=E_{m}|m\rangle$, the matrix elements of $ {\mathcal{T}}_{\alpha\alpha'}$ are given by 
\begin{equation}
\begin{split}
&\left\langle m\left| {\mathcal{T}}_{\alpha\alpha'}\right|n\right\rangle=\pi\delta\left(\varepsilon_{\alpha}-\varepsilon_{\alpha'}+E_{m}-E_{n}\right)\left\langle m\left| {T}_{\alpha\alpha'}\right|n\right\rangle-\\
&-iP\frac{1}{\varepsilon_{\alpha}-\varepsilon_{\alpha'}+E_{m}-E_{n}}\left\langle m\left| {T}_{\alpha\alpha'}\right|n\right\rangle.
\end{split}
\label{eqn:NewOpsElements}
\end{equation}
They include singularities at $\varepsilon_{\alpha}-\varepsilon_{\alpha'}=E_{n}-E_{m}$ which disappear after integrating over quasicontinuous spectra of electronic momentum in the leads, as we show below.

\subsubsection{Equation of motion for CSDMs}

Substituting Eq. (\ref{eqn:Cfinal}) into Eq. (\ref{eqn:EOMnonliouv}) results in the equation of motion for CSDMs 
\begin{equation}
\partial_{t} {\rho}_{n}=\mathcal{L} {\rho}_{n}-\mathcal{J} {\rho}'_{n}+\mathcal{D} {\rho}''_{n}
\label{eqn:MEforCSDM}
\end{equation}
(see Appendix \ref{sec:EOMforCSDM} for derivation), where $ {\rho}'_{n}$ and $ {\rho}''_{n}$ stand for the discrete derivatives 
\begin{equation}
\begin{split}
& {\rho}'_{n}=\frac{1}{2}\left( {\rho}_{n+1}- {\rho}_{n-1}\right),\\
& {\rho}''_{n}= {\rho}_{n+1}+ {\rho}_{n-1}-2 {\rho}_{n},
\end{split}
\label{eqn:DiscreteDerivatives}
\end{equation}
and $\mathcal{L}$, $\mathcal{J}$, $\mathcal{D}$ are linear superoperators defined below.
The superoperator $\mathcal{L}$ is responsible for the evolution of the density matrix.
Its action on a generic matrix $ {\chi}$ is given by 
\begin{equation}
\begin{split}
&\mathcal{L} {\chi}=-i\left[ {H}'_{A}, {\chi}\right]+\sum_{\eta\eta'\alpha\alpha'}J_{\eta}J_{\eta'}\left(1-f_{\alpha}\right)f_{\alpha'}\times\\
&\times\left( {\mathcal{T}}_{\alpha\alpha'} {\chi} {T}_{\alpha\alpha'}^{\dag}-\frac{1}{2}\left\{ {T}_{\alpha\alpha'}^{\dag} {\mathcal{T}}_{\alpha\alpha'}, {\chi}\right\}+\mbox{h.c.}\right),
\end{split}
\label{eqn:L-superoperatorDef}
\end{equation}
where the curly braces stand for the anticommutator and
\begin{equation}
 {H}'_{A}= {H}_{A}+\Delta {H}_{A},\label{eqn:Hshifted}
\end{equation}
accounts for the autonomous evolution of the atoms governed by the Hamiltonian $H_{A}$ and corrected by the coupling to the leads as 
\begin{equation}
\begin{split}
&\Delta {H}_{A}=\sum_{\eta\alpha}J_{\eta}f_{\alpha} {T}_{\alpha\alpha}+\sum_{\eta\eta'\alpha\alpha'}J_{\eta}J_{\eta'}\times\\
&\times\left(1-f_{\alpha}\right)f_{\alpha'}\frac{1}{2i}\left( {T}_{\alpha\alpha'}^{\dag} {\mathcal{T}}_{\alpha\alpha'}- {\mathcal{T}}_{\alpha\alpha'}^{\dag} {T}_{\alpha\alpha'}\right).
\end{split}
\label{eqn:HamilShift}
\end{equation}
The superoperators $\mathcal{J}$ and $\mathcal{D}$ acting on an arbitrary matrix $\chi$ are defined as
\begin{equation}
\begin{split}
&\mathcal{J} {\chi}=\mathcal{D}_{+} {\chi}-\mathcal{D}_{-} {\chi},\\
&\mathcal{D} {\chi}=\frac{1}{2}\left(\mathcal{D}_{+} {\chi}+\mathcal{D}_{-} {\chi}\right),
\end{split}
\label{eqn:DJ-definition}
\end{equation}
with 
\begin{equation}
\begin{split}
&\mathcal{D}_{+} {\chi}=J_{T}J_{S}\sum_{st}\left(1-f_{s}\right)f_{t}\left( {T}_{st} {\chi} {\mathcal{T}}_{st}^{\dag}+ {\mathcal{T}}_{st} {\chi} {T}_{st}^{\dag}\right),\\
&\mathcal{D}_{-} {\chi}=J_{T}J_{S}\sum_{st}\left(1-f_{t}\right)f_{s}\left( {T}_{ts} {\chi} {\mathcal{T}}_{ts}^{\dag}+ {\mathcal{T}}_{ts} {\chi} {T}_{ts}^{\dag}\right),
\end{split}
\label{eqn:D-superoperators}
\end{equation}
where indices $t$ and $s$ parametrize electronic states in the tip and the substrate correspondingly.

\subsubsection{Summation over bands}

We now perform the summation over $k,k'$ in Eqs. (\ref{eqn:L-superoperatorDef}), (\ref{eqn:HamilShift}), and (\ref{eqn:D-superoperators}) for the specific case in which the operators $ {T}_{\alpha\alpha'}$ do not depend on the momenta and the bandwidth of the reservoirs is much larger than other energy scales, i.e., $W\gg\Delta_{\eta},U,eV,1/\beta$.
We introduce the index $\lambda=(\eta,r,\sigma)$ that enumerates quantum numbers of the reservoirs other than momentum, so that $\alpha=(\lambda,k)$ and $ {T}_{\alpha\alpha'}= {T}_{\lambda\lambda'}$.
Using Eq. (\ref{eqn:NewOpsElements}), we evaluate the following sum 
\begin{equation}
\begin{split}
& \sum_{kk'}(1-f_{\alpha})f_{\alpha'} {\mathcal{T}}_{\alpha\alpha'}=\varrho_{\eta\sigma}\varrho_{\eta'\sigma'}\mathcal{V}_{\eta}\mathcal{V}_{\eta'}\left(\frac{\pi}{\beta}T'_{\lambda\lambda'}-\right.\\
&\left.-iW{T}_{\lambda\lambda'}\ln4+i\ln\frac{2\beta W}{\pi}\left(\left(\mu_{\eta}-\mu_{\eta'}\right)T_{\lambda\lambda'}+\left[H_{A},T_{\lambda\lambda'}\right]\right)\right)
\end{split}
\label{eqn:SumOverMomenta}
\end{equation}
(see Appendix \ref{sec:WBA} for derivation), where $T'_{\lambda\lambda'}$ are operators with matrix elements 
\begin{equation}
\left\langle m\left|T'_{\lambda\lambda'}\right|n\right\rangle=g\left(\beta\left(\mu_{\eta}-\mu_{\eta'}+E_{m}-E_{n}\right)\right)\left\langle m\left|T_{\lambda\lambda'}\right|n\right\rangle,
\end{equation}
and $g\left(x\right)=x\left(e^{x}-1\right)^{-1}$.
After substitution into Eq. (\ref{eqn:MEforCSDM}), the imaginary part of Eq. (\ref{eqn:SumOverMomenta}) contributes to the Hamiltonian shift (\ref{eqn:HamilShift}) as
\begin{equation}
\begin{split}
& \Delta {H}_{A}=\frac{W}{\pi}\sum_{\lambda}\gamma_{\lambda}T_{\lambda\lambda}+\frac{1}{\pi\beta}\sum_{\lambda\lambda'}\gamma_{\lambda}\gamma_{\lambda'}\times\\
& \times\frac{1}{2i}\left(T_{\lambda\lambda'}^{\dag}T'_{\lambda\lambda'}-\text{h.c.}\right)-\frac{W\ln4}{\pi^{2}}\sum_{\lambda\lambda'}\gamma_{\lambda}\gamma_{\lambda'}T_{\lambda\lambda'}^{\dag}T_{\lambda\lambda'}+\\
& +\frac{1}{2\pi^{2}}\ln\frac{2\beta W}{\pi}\left[H_{A},\sum_{\lambda\lambda'}\gamma_{\lambda}\gamma_{\lambda'}T^{\dag}_{\lambda\lambda'}T_{\lambda\lambda'}\right],
\end{split}
\label{eqn:HamilShiftSummed}
\end{equation}
where we have identified the parameters $\gamma_{\lambda}=\pi J_{\eta}\varrho_{\eta\sigma}\mathcal{V}_{\eta}$.
Substituting Eq. (\ref{eqn:SumOverMomenta}) into Eq. (\ref{eqn:L-superoperatorDef}), one obtains 
\begin{equation}
\begin{split}
& \mathcal{L} {\chi}=-i\left[H'_{A},\chi\right]+\frac{1}{\pi\beta}\sum_{\lambda\lambda'}\gamma_{\lambda}\gamma_{\lambda'}\times\\
& \times\left(T''_{\lambda\lambda'}\chi T_{\lambda\lambda'}^{\dag}-\frac{1}{2}\left\{T_{\lambda\lambda'}^{\dag}T''_{\lambda\lambda'},\chi\right\} +\text{h.c.}\right).
\end{split}
\label{eqn:LsuperSummed}
\end{equation}
where we defined
\begin{equation}
T''_{\lambda\lambda'}=T'_{\lambda\lambda'}+i~\frac{\beta}{\pi}\ln\frac{2\beta W}{\pi}\left[H_{A},T_{\lambda\lambda'}\right].
\label{eqn:TppOperators}
\end{equation}
In a similar way Eq. (\ref{eqn:D-superoperators}) becomes
\begin{equation}
\begin{split}
& \mathcal{D}_{+} {\chi}=\frac{1}{\pi\beta}\sum_{\lambda_{S}\lambda_{T}}\gamma_{\lambda_{S}}\gamma_{\lambda_{T}}\left(T''_{\lambda_{S}\lambda_{T}}\chi T_{\lambda_{S}\lambda_{T}}^{\dag}+\text{h.c.}\right),\\
& \mathcal{D}_{-} {\chi}=\frac{1}{\pi\beta}\sum_{\lambda_{S}\lambda_{T}}\gamma_{\lambda_{T}}\gamma_{\lambda_{S}}\left(T''_{\lambda_{T}\lambda_{S}}\chi T_{\lambda_{T}\lambda_{S}}^{\dag}+\text{h.c.}\right).
\end{split}
\label{eqn:DsuperSummed}
\end{equation}
In the following we do not take the imaginary part of the operators (\ref{eqn:TppOperators}) into account, as it leads to unphysical results.
We believe that this term is an artifact of performed approximations and would vanish in a more rigorous treatment, e.g., going beyond the Born approximation.
We thus use $T'_{\lambda\lambda'}$ instead of $T''_{\lambda\lambda'}$ in Eqs. (\ref{eqn:LsuperSummed}) and (\ref{eqn:DsuperSummed}).
We however leave the corresponding logarithmic term in the Hamiltonian shift (\ref{eqn:HamilShiftSummed}), as it has a physical meaning \cite{Oberg2014}.

\subsubsection{Master equation}

As stated in Sec. \ref{sec:MethodSummary}, $\mathcal{L}$ determines the evolution of the atomic subsystem.
This can be seen by summing Eq. (\ref{eqn:MEforCSDM}) over charge-specific components which leads to the equation of motion for the unconditioned density matrix $ {\rho}=\sum_{n} {\rho}_{n}$.
We use $\sum_{n} {\rho}'_{n}=0$ and $\sum_{n} {\rho}''_{n}=0$ to obtain
\begin{equation}
\partial_{t} {\rho}=\mathcal{L} {\rho}.
\label{eqn:MEforDM}
\end{equation}
In principle, this equation can be put in a canonical form in order to identify the coherence rates that characterize the dissipative dynamics \cite{Hall2014}.
We were not able to perform this procedure in general but observed in the specific examples below that the decoherence rates
are not always positive.
This implies that the evolution of the density matrix is generally non-Markovian.
A general proof that the density matrix evolving according to Eq. (\ref{eqn:MEforDM}) remains positively defined has also not been found.
Nevertheless, for all the examples worked out in Sec. \ref{sec:Results} we checked numerically that this was the case.
We note that the usual Markovian master equation is recovered in some limiting cases; see Sec. \ref{sec:Application}.

\subsubsection{Current}

The probability that $n$ electrons have been transferred from the tip to the substrate is given by $p_{n}=\mbox{tr} {\rho}_{n}$.
The average current from the tip to the substrate is thus given by $I=-e\partial_{t}\langle n\rangle=-e\mbox{tr}\sum_{n}n\partial_{t} {\rho}_{n}$.
Using Eq. (\ref{eqn:MEforCSDM}) and relations $\sum_{n}n {\rho}'_{n}=- {\rho}$, $\sum_{n}n {\rho}''_{n}=0$, one can show that 
\begin{equation}
I=-e\,\mbox{tr}\mathcal{J} {\rho}.
\label{eqn:Current}
\end{equation}
The steady state value of the current is calculated by substituting $ {\rho}= {\rho}_{\infty}$ into Eq. (\ref{eqn:Current}), where the steady state density matrix $\rho_{\infty}$ is calculated as the eigenstate of $\mathcal{L}$ associated with zero eigenvalue.

\subsubsection{Shot noise}

Fluctuations of the current are characterized by the shot noise defined as 
\begin{equation}
S=2e^{2}\partial_{t}\left(\langle n^{2}\rangle-\langle n\rangle^{2}\right).
\end{equation}
Using arguments similar to those for the current, one can show that 
\begin{equation}
S=4e^{2}\mbox{tr}\left(\mathcal{D} {\rho}+\mathcal{J}\sum_{n}\left(n-\langle n\rangle\right)\rho_{n}\right).
\label{eqn:Shotnoise}
\end{equation}
In contrast to the case of average current, the shot noise cannot be expressed through the density matrix alone.
One also needs to evaluate the quantity $\sum_{n}(n-\langle n\rangle) {\rho}_{n}= {\rho}^{(1)}$ which satisfies the equation of motion 
\begin{equation}
\partial_{t} {\rho}^{(1)}=\mathcal{L} {\rho}^{(1)}+\mathcal{J} {\rho}- {\rho}\mbox{tr}\mathcal{J} {\rho}.
\label{eqn:MEforDM1}
\end{equation}
In the steady state we obtain 
\begin{equation}
\mathcal{L} {\rho}_{\infty}^{(1)}= {\rho}_{\infty}\mbox{tr}\left(\mathcal{J} {\rho}_{\infty}\right)-\mathcal{J} {\rho}_{\infty},
\label{MEforDM1_2}
\end{equation}
which has a formal solution
\begin{equation}
\rho_{\infty}^{(1)}=-\mathcal{L}^{-1}\mathcal{J}\rho_{\infty},
\label{eqn:rho1steady}
\end{equation}
(see Appendix \ref{sec:Inversion}), where $\mathcal{L}^{-1}$ is the pseudoinverse of $\mathcal{L}$,
i.e., taken excluding the zero eigenvalue of $\mathcal{L}$.

\subsection{Application to EASD}
\label{sec:Application}

Here we apply the presented method to the model of the EASD introduced in Sec. \ref{sec:Model}.
In particular, we specify Eqs. (\ref{eqn:HamilShiftSummed}), (\ref{eqn:LsuperSummed}), (\ref{eqn:DsuperSummed}) using the coupling Hamiltonian (\ref{eqn:Coupling}) that may be recovered from the generic one used in Sec. \ref{sec:MethodDetails} by the substitution
\begin{equation}
T_{\lambda\lambda'}=S_{r\sigma\sigma'}\delta_{rr'}\left(\delta_{rr_{0}}+\left(1-\delta_{rr_{0}}\right)\delta_{\eta S}\delta_{\eta'S}\right),
\label{eqn:EASDjumpopers}
\end{equation}
where $\lambda=\left(\eta,r,\sigma\right)$ and $ {S}_{r\sigma\sigma'}$ stands for the atomic operators
\begin{equation}
 {S}_{r\sigma\sigma'}=
\begin{cases}
~ {S}_{rz} & \mbox{if}~\sigma=\sigma'=\uparrow,\\
~ {S}_{r+}= {S}_{rx}+i {S}_{ry} & \mbox{if}~\sigma=\downarrow,~\sigma'=\uparrow,\\
~ {S}_{r-}= {S}_{rx}-i {S}_{ry} & \mbox{if}~\sigma=\uparrow,~\sigma'=\downarrow,\\
~- {S}_{rz} & \mbox{if}~\sigma=\sigma'=\downarrow.
\end{cases}
\label{eqn:Soperators}
\end{equation}
The delta functions are introduced in Eq. (\ref{eqn:EASDjumpopers}) to account for the features of the model: (i) electrons only tunnel between the leads coupled to the same atom; (ii) the tip is only coupled to the atom $r_{0}$.

As shown in Appendix \ref{sec:EASD}, the resulting expressions for Eqs. (\ref{eqn:HamilShiftSummed}), (\ref{eqn:LsuperSummed}), (\ref{eqn:DsuperSummed}) include $ {S}_{r\sigma\sigma'}$ and operators $Q_{r\sigma\sigma'}^{(0)}$, $Q_{r\sigma\sigma'}^{(+)}$ and $Q_{r\sigma\sigma'}^{(-)}$ whose matrix elements are given by
\begin{equation}
\begin{split}
&\langle m|Q_{r\sigma\sigma'}^{(0)}|n\rangle=g\left(\beta\left(E_{m}-E_{n}\right)\right)\langle m| {S}_{r\sigma\sigma'}|n\rangle,\\
&\langle m|Q_{r\sigma\sigma'}^{(+)}|n\rangle=g\left(\beta\left(E_{m}-E_{n}+eV\right)\right)\langle m| {S}_{r\sigma\sigma'}|n\rangle,\\
&\langle m|Q_{r\sigma\sigma'}^{(-)}|n\rangle=g\left(\beta\left(E_{m}-E_{n}-eV\right)\right)\langle m| {S}_{r\sigma\sigma'}|n\rangle.
\end{split}
\label{eqn:Qoperators}
\end{equation}
For the Hamiltonian shift (\ref{eqn:HamilShiftSummed}) we obtain 
\begin{equation}
\begin{split}
& \Delta H_{A}=\frac{1}{\pi\beta}\sum_{r\sigma\sigma'}\frac{1}{2i}\left(S_{r\sigma\sigma'}^{\dag}A_{r\sigma\sigma'}-\text{h.c.}\right)+\frac{1}{\pi}p\gamma_{T}W S_{r_{0}z}- \\
& -\frac{8\ln2}{\pi^{2}}p^{2}\gamma_{T}^{2}W S_{r_{0}z}^{2}+\frac{2}{\pi^{2}}p^{2}\gamma^{2}_{T}\ln\frac{2\beta W}{\pi}\left[H_{A},S^{2}_{r_{0}z}\right]+C,
\end{split}
\label{eqn:HamilShiftEASD}
\end{equation}
where the constant part is given by 
\begin{equation}
\begin{split}
& C=\frac{4W\ln2}{\pi^{2}}\left(\gamma_{S}^{2}\sum_{r}\mathbf{S}_{r}^{2}+\right.\\
& \left.+2\gamma_{S}\gamma_{T}\mathbf{S}_{r_{0}}^{2}+\gamma_{T}^{2}(1-p^{2})\mathbf{S}_{r_{0}}^{2}\right),
\end{split}
\end{equation}
and we have introduced operators
\begin{equation}
\begin{split}
& A_{r\sigma\sigma'}=\delta_{rr_{0}}\gamma_{S}\gamma_{T}\left(w_{\sigma}Q_{r\sigma\sigma'}^{(+)}+w_{\sigma'}Q_{r\sigma\sigma'}^{(-)}\right)+ \\
& +\left(\gamma_{S}^{2}+\delta_{rr_{0}}\gamma_{T}^{2}w_{\sigma}w_{\sigma'}\right)Q_{r\sigma\sigma'}^{(0)}.
\end{split}
\label{eqn:Aoperators}
\end{equation}
The terms in the shift (\ref{eqn:HamilShiftEASD}), except for the first one, act as a renormalization of the magnetic field and the anisotropy parameters in Eq. (\ref{eqn:Cluster}).
We thus do not explicitly account for them in the numerical calculations.
For Eq. (\ref{eqn:LsuperSummed}) we obtain 
\begin{equation}
\begin{split}
&\mathcal{L} {\chi}=-i\left[ {H}'_{A}, {\chi}\right]+\frac{1}{\pi\beta}\sum_{r\sigma\sigma'}\left( {A}_{r\sigma\sigma'} {\chi} {S}_{r\sigma\sigma'}^{\dag}-\right.\\
&\left.-\frac{1}{2}\left\{  {S}_{r\sigma\sigma'}^{\dag} {A}_{r\sigma\sigma'}, {\chi}\right\} +\text{h.c.}\right).
\end{split}
\label{eqn:LsuperEASD}
\end{equation}
Finally, the result for Eq. (\ref{eqn:DsuperSummed}) is expressed as 
\begin{equation}
\begin{split}
&\mathcal{D}_{+} {\chi}=\frac{\gamma_{T}\gamma_{S}}{\pi\beta}\sum_{\sigma\sigma'}w_{\sigma'}\left(Q_{r_{0}\sigma\sigma'}^{(+)}\chi S_{r_{0}\sigma\sigma'}^{\dag}+\mbox{h.c.}\right),\\
&\mathcal{D}_{-} {\chi}=\frac{\gamma_{S}\gamma_{T}}{\pi\beta}\sum_{\sigma\sigma'}w_{\sigma}\left(Q_{r_{0}\sigma\sigma'}^{(-)}\chi S_{r_{0}\sigma\sigma'}^{\dag}+\mbox{h.c.}\right).
\end{split}
\label{eqn:DsuperEASD}
\end{equation}
The superoperator (\ref{eqn:LsuperEASD}) of the master equation has the Lindblad form when $ {A}_{r\sigma\sigma'}\sim {S}_{r\sigma\sigma'}$.
As shown in Appendix \ref{sec:Lindblad}, this happens in the following cases: (i) infinite temperature $\beta\to\infty$, (ii) infinite voltage $|V|\to\infty$, (iii) single atom in the parallel magnetic field $\mathbf{B}\parallel\mathbf{P}$.
The obtained superoperator does not couple the diagonal and off-diagonal elements of the density matrix in the case of a single atom and $\left[H_{A},S_{z}\right]=0$.
We thus always get the equivalent results with the method of rate equations for single atoms in the parallel geometry, as shown in the next section.

\section{Results}
\label{sec:Results}

In this section we provide two examples using the equations derived above: (i) a single spin in the presence of a spin-polarized tip, and (ii) a spin chain.
In addition to the transport properties and observables of the atomic subsystem, we also compute the von Neumann entropy $S=-\mbox{tr}\left(\rho\ln\rho\right)$ that characterizes the degree of purity of the atomic state.

\subsection{Single atom with $S=1/2$}
\label{subsec:Single1/2}

The simplest example of a magnetic structure is an atom with spin $S=1/2$ for which the density matrix can be expressed through the average spin projections as $\rho=\frac{1}{2}+\langle\mathbf{S}\rangle\cdot\mathbf{\tau}$.
In this case the anisotropy terms in the Hamiltonian may be discarded as they only yield a constant energy contribution.
The Hamiltonian is thus reduced to the contribution of the external magnetic field $\mathbf{B}$ yielding a Zeeman energy gap $\Delta=g\mu_{B}\abs{\mathbf{B}}$ between two energy levels of the atom.

For $V=0$ relaxation processes due to interaction with the electronic leads bring the atom to a thermal state $ {\rho}\propto e^{-\beta {H}_{A}}$. At low temperatures $\beta>\Delta^{-1}$ the atomic spin is fully polarized along the magnetic field.
A finite applied voltage  $V \neq 0$ causes current to ensue through the atom, inducing spin excitations and changing the atomic steady state.
The inelastic contribution to the current results from a spin-flip process $|\uparrow\rangle\rightarrow|\downarrow\rangle$ driven by tunneling electrons.

In the following we choose the parameters $g=2$, $B=5$ T, $\gamma_{T}=\gamma_{S}=0.8$, $(\beta k_{B})^{-1}=1$ K and vary the value of the polarization  $p=0, 0.5, 1$. 
These are typical experimental parameters \cite{Hirjibehedin2006,Hirjibehedin2007} within the applicability domain of our method.

We investigate the steady state of the atom, the differential conductance $dI/dV$ and the differential shot noise $dS/dV$.
In order to identify the contribution due to coherences, we compare our results, obtained with master equation (ME), with those obtained using rate equations (REs).
As explained above, the ME method deals with the full density matrix and thus accounts for coherence effects, in contrast to REs that operate with the diagonal elements of $\rho$.
However, the master equation cannot be used to study nonperturbative phenomena, such as Kondo correlations, unless the lead-atom coupling is treated beyond the second order.
We consider two different geometries where the applied field is either parallel or perpendicular to the polarization vector of the tip.

In the parallel geometry, when both $\mathbf{B}$ and $\mathbf{P}$ are along the $z$ axis, the ME and REs yield equivalent spectra for any polarization parameter.
Indeed, since $\langle S_{x}\rangle=\langle S_{y}\rangle=0$, off-diagonal elements of the density matrix vanish and coherences do not affect the average current and the shot noise.
The curves for the steady state observables are given in Appendix \ref{sec:Parallel} and reproduce already known results \cite{Delgado2010,Delgado2010a}.

\begin{figure}[t]
\includegraphics[width=1\columnwidth]{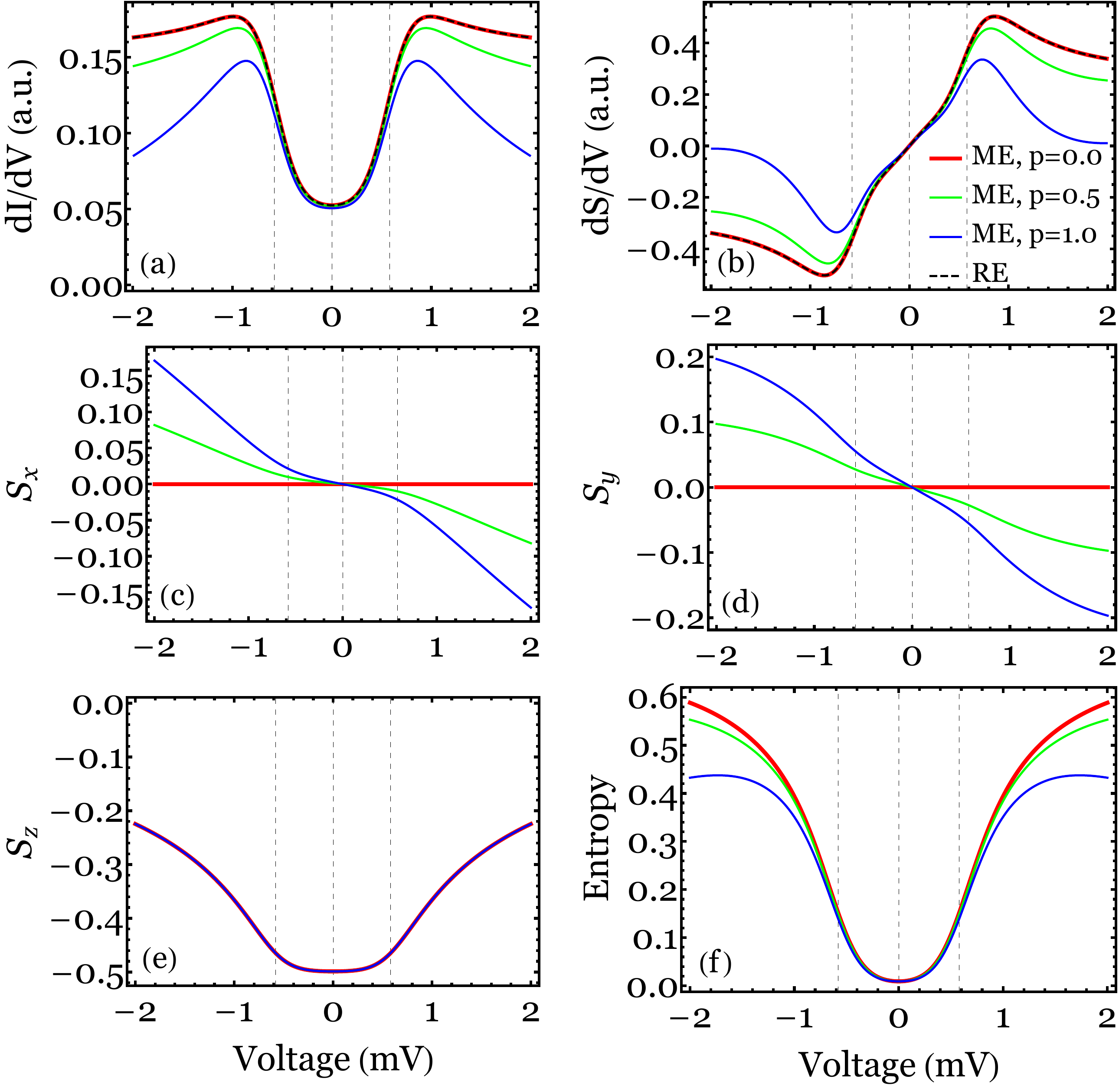}
\caption{Steady state characteristics for a single spin $S=1/2$ in a perpendicular geometry ($\mathbf{B}$ along $z$ axis, $\mathbf{P}$ along $x$ axis) for different values of the polarization parameter $p$ and for $g=2$, $B=5$ T, $\gamma_{T}=\gamma_{S}=0.8$, $(\beta k_{B})^{-1}=1$ K.
The quantities presented as functions of voltage are (a) differential conductance, (b) differential shot noise, (c) average spin component $\langle S_{x}\rangle$, (d) average spin component $\langle S_{y}\rangle$, (e) average spin component $\langle S_{z}\rangle$, (f) entropy.
In this geometry RE and ME approaches are not equivalent for $p\neq0$, as the coherences determined by $\langle S_{x}\rangle$ and $\langle S_{y}\rangle$ do not vanish and give contribution to the results.
While ME gives different $dI/dV$ and $dS/dV$ curves for different $p$, the results obtained with REs are independent of $p$ and coincides with the ME results for $p=0$.}
\label{fgr:SingleOneHalfPerpendicular} 
\end{figure}

The calculated spectra in the perpendicular geometry, when $\mathbf{B}$ is along the $z$ axis and $\mathbf{P}$ is along the $x$ axis, are shown in Fig. \ref{fgr:SingleOneHalfPerpendicular}.
In this case the RE approach gives the same result for any $p$.
This is due to the fact that a change in the polarization parameter does not affect the spin population of electrons in the tip measured in a perpendicular direction.
Therefore, if coherences are ignored, a polarization perpendicular to the magnetic field applied to the spin should not affect the current.
On the contrary, if coherences are taken into account, the mismatch between the polarization of the electrons and the direction of the atomic spin reduces both the average current and the shot noise.
This decrease depends on the polarization parameter, reaching a maximum for $p=1$ (fully polarized tip) and vanishing for $p=0$ (unpolarized tip).
The clear difference between curves calculated with REs and the ME shows that in this geometry it is essential to take into account effects of coherences to correctly describe the average current and the shot noise.
In other words, interference effects within the atomic subsystem substantially modify its conductance properties.
It is worth noting that, although the spin is polarized in the $z$ direction and the magnetic field is in the $x$ direction, all three components of the spin acquire a nonzero mean value.
This effect is a direct result of a spin transfer torque \cite{Slonczewski1996}.
It has been studied theoretically in quantum dots coupled to magnetic leads in noncollinear arrangements \cite{Konig2003,Braun2004,Rudzinski2004,Weymann2007}.
For larger voltages we observe that the entropy is suppressed as the polarization degree of the tip is increased.

\begin{figure}[t]
\includegraphics[width=1\columnwidth]{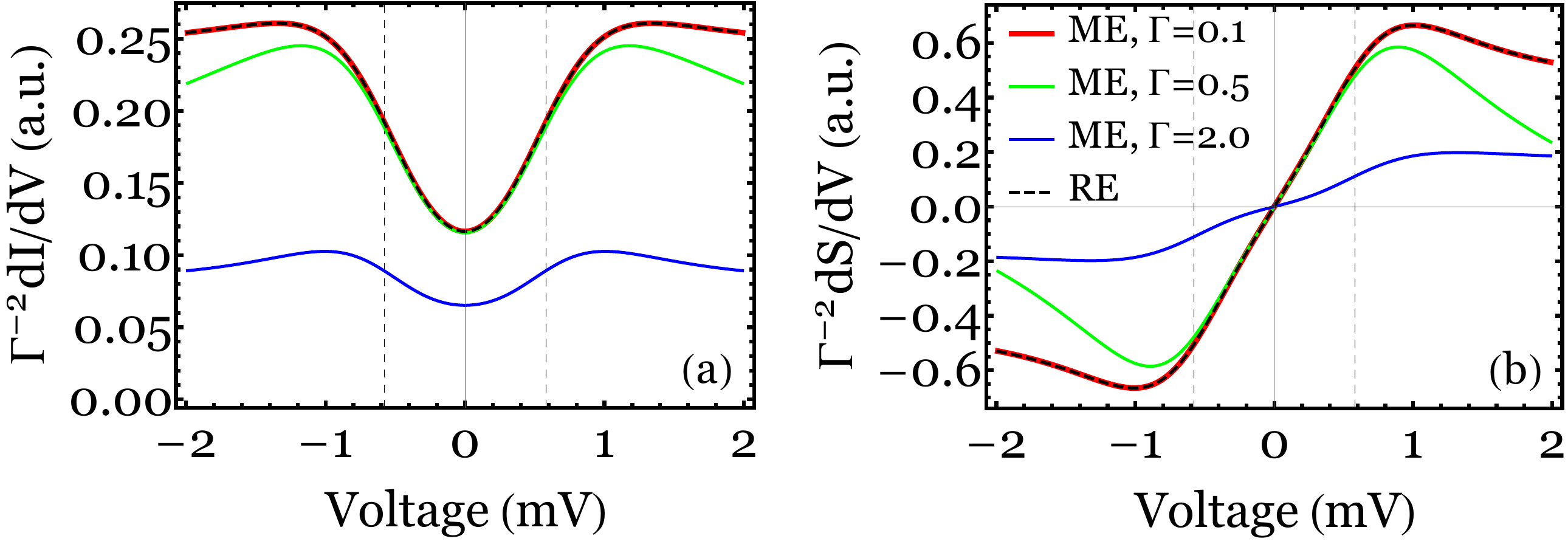}
\caption{Dependence of (a) average current and (b) shot noise on the coupling strength for a single spin $S=1/2$ in a perpendicular geometry ($\mathbf{B}$ along $z$ axis, $\mathbf{P}$ along $x$ axis).
$dI/dV$ and $dS/dV$ curves are computed with REs and ME for different values of $\gamma_{T}=\gamma_{S}=\gamma$ and for $g=2$, $B=5$ T, $(\beta k_{B})^{-1}=2$ K, $p=1$.
In the limit of weak coupling $dI/dV$ and $dS/dV$ curves obtained with ME coincide with the results obtained with REs.}
\label{fgr:SingleOneHalfPerpendicularCoupling} 
\end{figure}

To analyze the dependence of the inelastic current on the coupling strength, in Fig. \ref{fgr:SingleOneHalfPerpendicularCoupling} we compare $dI/dV$ and $dS/dV$ curves scaled by a $\gamma^{-2}$ factor for different values of $\gamma=\gamma_{S}=\gamma_{T}$.
As expected, for a vanishing coupling both RE and ME methods yield the same results since the relative contribution of coherences to $dI/dV$ and $dS/dV$ vanishes.
To emphasize this contribution to the spectra and make it more pronounced, we use the values of $\gamma$ at the limit of validity of the Born approximation.

\subsection{Single atom with $S=5/2$}

Atoms used in spin-polarized STM experiments typically have spins higher than $S=1/2$.
Therefore we now analyze the case of a Mn atom with spin $S=5/2$.
Here, even in the absence of external magnetic field, the energy levels can be split by the anisotropy terms.
For $D<0$ the states with $S_{z}=+5/2$ and $S_{z}=-5/2$ are separated by the energy barrier and may be used for quantum information storage \cite{Miyamachi2013}.
In the following we set $D=-0.04$ meV, $E=0$, $g=2$, $B=0$ T, $\gamma_{S}=\gamma_{T}=0.6$ and $(\beta k_{B})^{-1}=0.5$ K, taken from Refs. \cite{Hirjibehedin2006,Hirjibehedin2007}.
We do not consider the case $E\neq0$ separately, as the corresponding results are not qualitatively different from the ones presented below for the perpendicular geometry.
The transport through nanomagnets has been previously studied in a number of papers \cite{Timm2006,Elste2006,Elste2007,Misiorny2007,Misiorny2009}.
Here, we focus on the difference between the results of the ME method that takes into account coherences and the ones obtained within the previous approaches based on the rate equations.

In the parallel geometry, with both $\mathbf{B}$ and $\mathbf{P}$ along the $z$ axis, the ME and RE approaches give the same results, similarly to the single atom with spin $S=1/2$.
The spectra of the steady state observables are shown in Appendix \ref{sec:Parallel} and coincide with ones presented in Refs. \cite{Delgado2010,Delgado2010a}.

\begin{figure}[t]
\includegraphics[width=1\columnwidth]{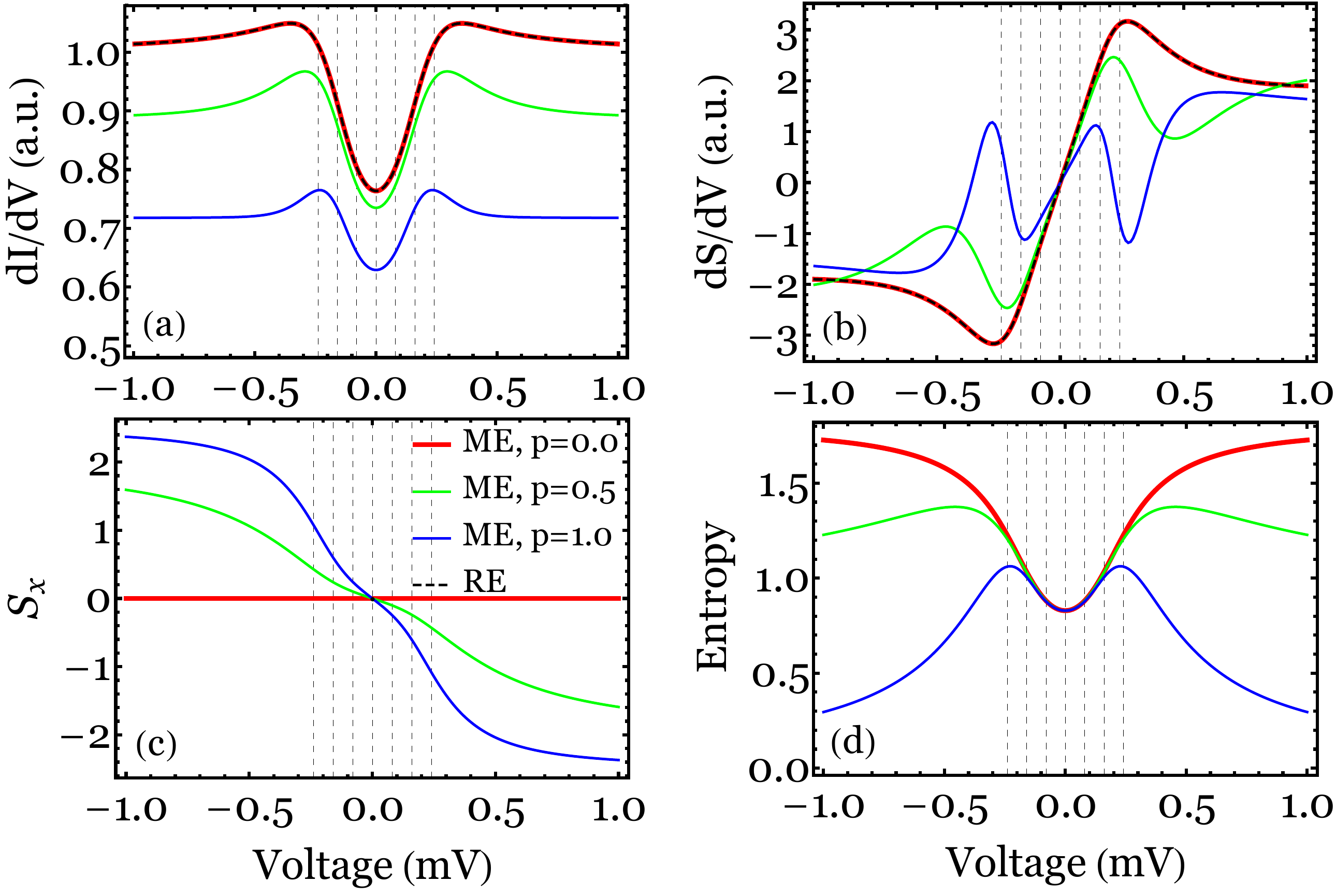}
\caption{Steady state characteristics for a single spin $S=5/2$ in a perpendicular geometry ($z$ is the easy axis of the crystal and $\mathbf{P}$ is along $x$ axis) for different values of the polarization parameter $p$ and for $D=-0.04$ meV, $\gamma_{T}=\gamma_{S}=0.6$, $(\beta k_{B})^{-1}=0.5$ K.
The quantities presented as functions of voltage are (a) differential conductance, (b) differential shot noise, (c) average spin component $\langle S_{x}\rangle$, (d) entropy.
In this geometry RE and ME approaches are not equivalent for $p\neq0$, as the coherences determined by $\langle S_{x}\rangle$ do not vanish and give contribution to the results.
Other components of the spin vanish, i.e., $\langle S_{y}\rangle=\langle S_{z}\rangle=0$.
While ME gives different $dI/dV$ and $dS/dV$ curves for different $p$, the results obtained with REs are independent of $p$ and coincides with the ME results for $p=0$.}
\label{fgr:SingleFiveHalfPerpendicular}
\end{figure}

The spectra of the steady state current in the perpendicular geometry, when $z$ is the easy axis and $\mathbf{P}$ is along $x$ axis, are shown in Fig. \ref{fgr:SingleFiveHalfPerpendicular}.
In this case the RE approach gives slightly different curves for different $p$, in contrast to the single atom with spin $S=1/2$. However, we do not show this difference as it is small compared to the contribution due to coherences that grows with the polarization parameter.
The switching of the atom to the state whose magnetization is collinear with the tip polarization requires higher voltages than for the parallel geometry.
That is explained by the change in the atomic spectrum due to the magnetic field produced by the polarized current.
The switching occurs for the polarized tip with $p\neq0$ and is accompanied by the decrease in the entropy as the voltage goes up.
For the unpolarized tip $p=0$, there is no switching and the entropy monotonically increases with the voltage.

\subsection{Spin-$1/2$ chain}

\begin{figure}[t]
\includegraphics[width=1\columnwidth]{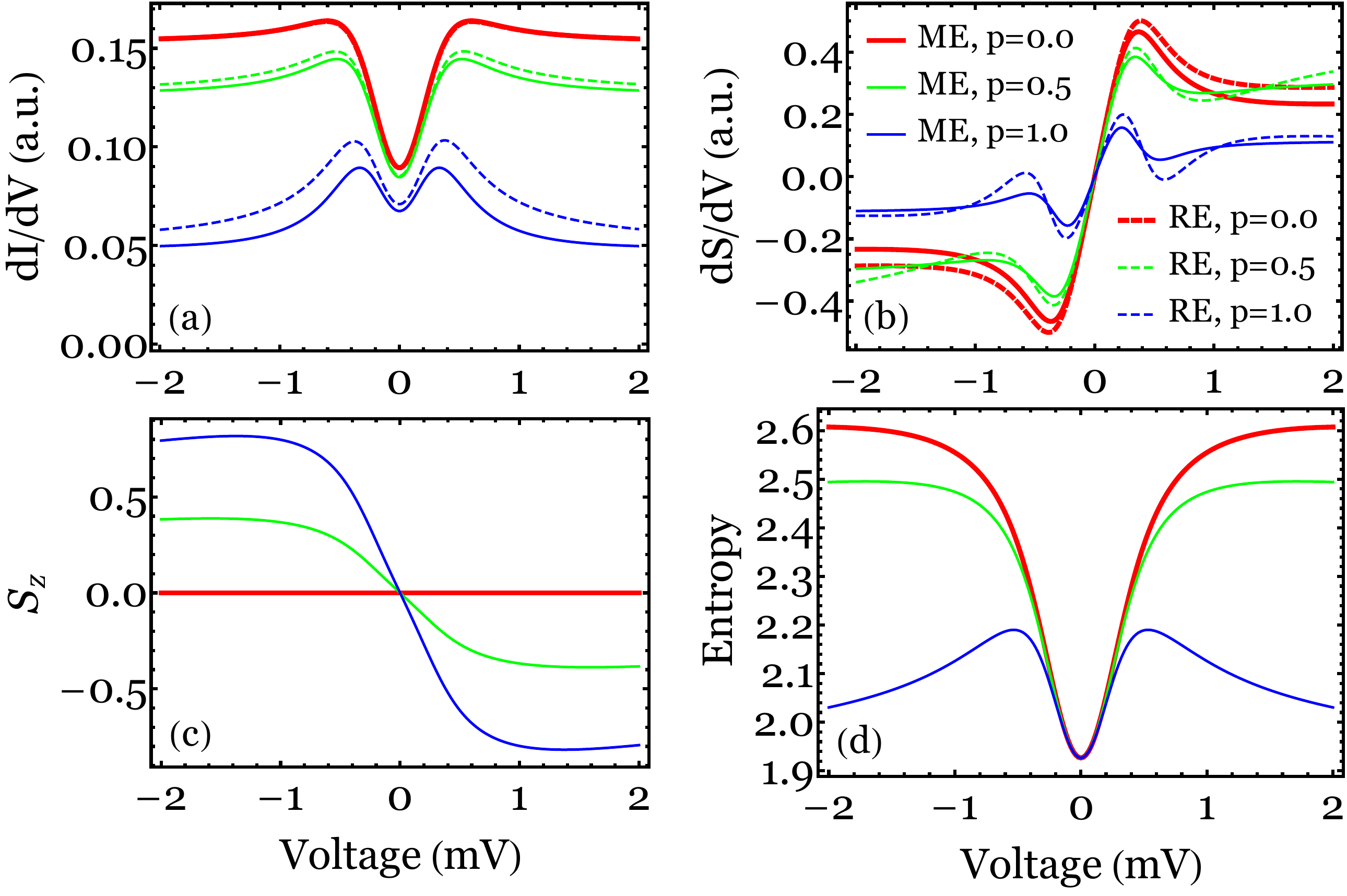}
\caption{Steady state characteristics for a chain of 4 spins $S=1/2$ in zero magnetic field for different values of the polarization parameter $p$ and for $(\beta k_{B})^{-1}=1$ K, $\gamma_{T}=\gamma_{S}=0.8$, $J=0.3$ meV.
The quantities presented as functions of voltage are (a) differential conductance, (b) differential shot noise, (c) average spin component $\langle S_{z}\rangle$, (d) entropy.
Other components of the spin vanish, i.e., $\langle S_{x}\rangle=\langle S_{y}\rangle=0$.}
\label{fgr:ChainZero} 
\end{figure}

\begin{figure}[t]
\includegraphics[width=1\columnwidth]{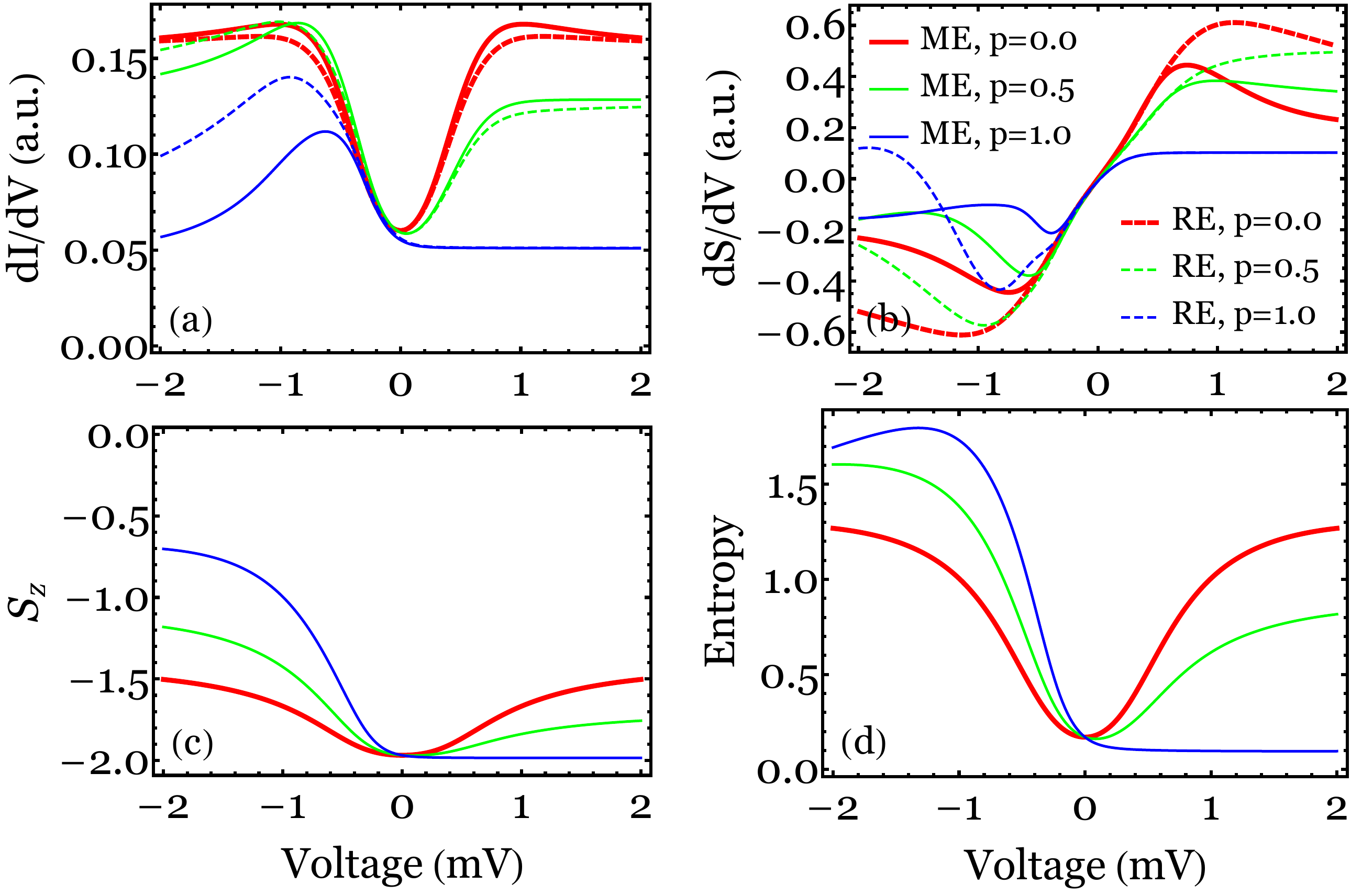}
\caption{Steady state characteristics for a chain of 4 spins $S=1/2$ in a parallel geometry (both $\mathbf{B}$ and $\mathbf{P}$ along $z$ axis) for different values of the polarization parameter $p$ and for $g=2$, $B=5$ T, $(\beta k_{B})^{-1}=1$ K, $\gamma_{T}=\gamma_{S}=0.8$, $J=0.3$ meV.
The quantities presented as functions of voltage are (a) differential conductance, (b) differential shot noise, (c) average spin component $\langle S_{z}\rangle$, (d) entropy.
Other components of the spin vanish, i.e., $\langle S_{y}\rangle=\langle S_{z}\rangle=0$.}
\label{fgr:ChainParallel} 
\end{figure}

\begin{figure}[t]
\includegraphics[width=1\columnwidth]{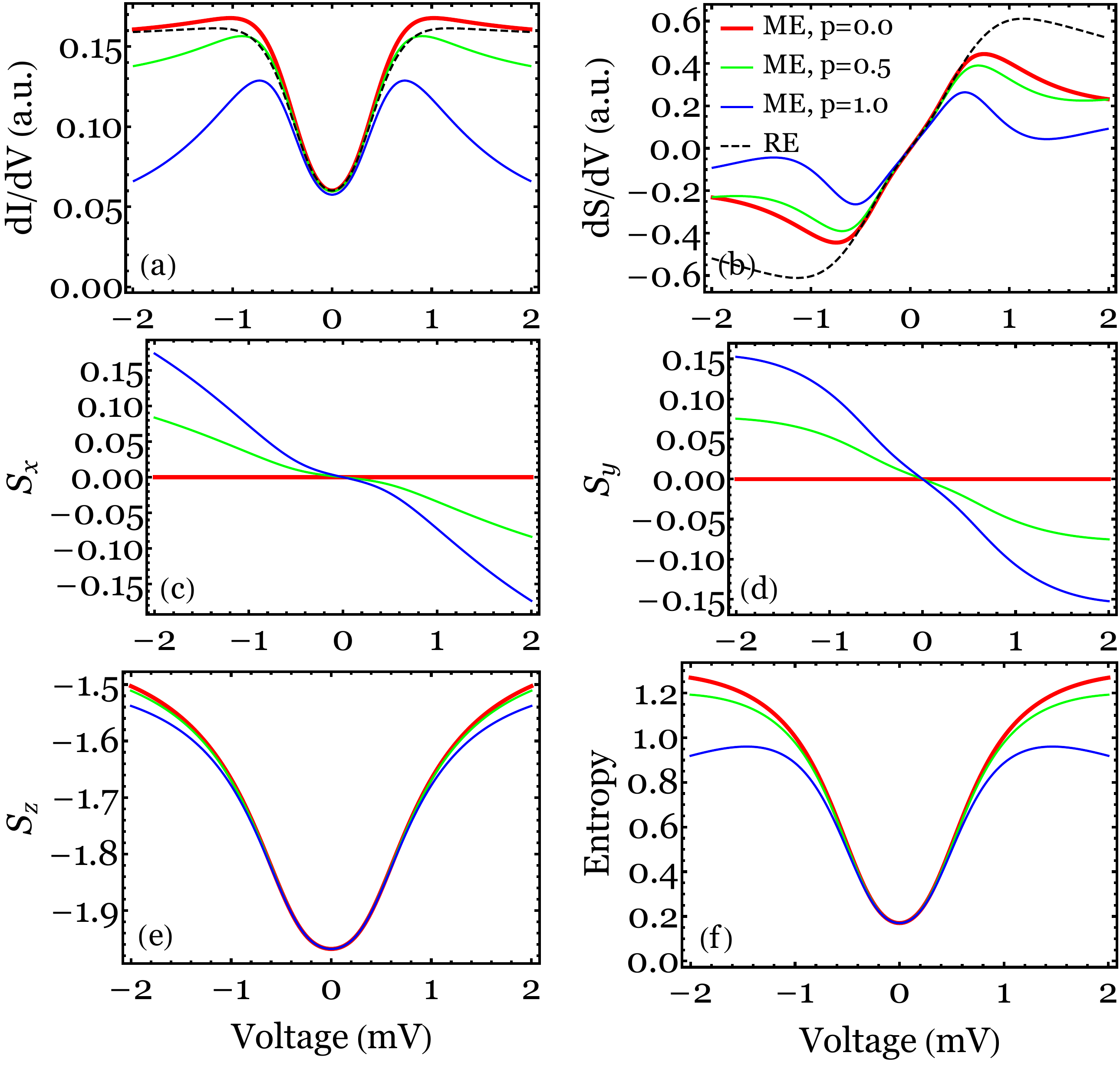}
\caption{Steady state characteristics for a chain of 4 spins $S=1/2$ in a perpendicular geometry ($\mathbf{B}$ along $z$ axis, $\mathbf{P}$ along $x$ axis) for different values of the polarization parameter $p$ and for $g=2$, $B=5$ T, $(\beta k_{B})^{-1}=1$ K, $\gamma_{T}=\gamma_{S}=0.8$, $J=0.3$ meV.
The quantities presented as functions of voltage are (a) differential conductance, (b) differential shot noise, (c) average spin component $\langle S_{x}\rangle$, (d) average spin component $\langle S_{y}\rangle$, (e) average spin component $\langle S_{z}\rangle$, (f) entropy.}
\label{fgr:ChainPerpendicular} 
\end{figure}

The manipulation capabilities of STM can be used to assemble chains of magnetic atoms on the substrate.
Compared to the case of single atoms, the conductivity profile of an atom in the chain is modified by the inter-atomic coupling.
Here we study the effect of coherences in the inelastic current when the tip drives a current through one of the atoms of a linear chain of $4$ atoms.
We consider the chain in the external magnetic field $B$ and study three geometries of the setup: (i) $\mathbf{B}=0$, (ii) $\mathbf{B}\neq0$, $\mathbf{B}\perp\mathbf{P}$, (iii) $\mathbf{B}\neq0$, $\mathbf{B}\parallel\mathbf{P}$.
The results calculated with the ME and RE methods are shown in Figs. \ref{fgr:ChainZero}, \ref{fgr:ChainParallel}, and \ref{fgr:ChainPerpendicular} for the same parameters as in Sec. \ref{subsec:Single1/2} and for the case when the tip is coupled to one of the central atoms $r=2$.

The spectra of the steady state current through the chain in zero magnetic field is presented in Fig. \ref{fgr:ChainZero}.
In this case the energy scale is set by the coupling constant $J=0.3$ meV.
Due to the antiferromagnetic coupling, the ground state of the chain has the total spin $S_{\text{tot}}=0$.
The difference between ME and RE approaches increases with $p$ for the $dI/dV$ curve and has the same order for all $p$ for the $dS/dV$ curve.
Driving the polarized current through the chain results in the switching to the collinearly polarized state, i.e., the state with the ferromagnetic order of spins.
The switching is accompanied by the decrease in the entropy as the voltage goes up.

In the case of parallel geometry, with both $\mathbf{B}$ and $\mathbf{P}$ along the $z$ axis, the two approaches give different results for any polarization parameter, including the unpolarized tip with $p=0$; i.e., the coherences contribute to the current.
That is in contrast to the case of a single spin, see Appendix \ref{sec:Parallel}, where coherences vanish.
The contribution of coherences is particularly noticeable in the shot noise which gets suppressed.
We explain this by the fact that the coupling drives individual atoms into a coherent superposition of states.
The entropy is smaller compared to the case of zero magnetic field.

In the case of perpendicular geometry, with $\mathbf{B}$ along the $z$ axis and $\mathbf{P}$ along the $x$ axis, the results obtained within two approaches are not equivalent for any polarization parameter, including $p=0$, differently from the case of a single atom, where the ME and RE results coincide for the unpolarized tip.
The difference between methods is especially remarkable for the shot noise calculations.
Note also that, similarly to the case of a single atom, see Fig. \ref{fgr:SingleOneHalfPerpendicular}, the RE approach yields the same result for different tip polarizations.

\section{Conclusion}
\label{sec:Discussion}

A master equation of the Redfield type describing the dynamics of the density matrix of an atomic spin structure was derived in the limit of a small lead-atom coupling and a short lead memory time, as compared with the energy and time scales of the isolated atomic spin system.
Its generalization to charge-specific density matrices allows for the description of transport quantities such as the current and the shot noise, in addition to the observables of the atomic subsystem. 

Unlike approaches based on rate equations, this description accounts for the dynamics of coherences, i.e., the off-diagonal elements of the density matrix.
It is suitable to describe the moderate lead-atom coupling regime where coherences cannot be disregarded.
This approach is however unable to capture nonperturbative phenomena in the lead-atom coupling such as Kondo effect and may yield unphysical results for large coupling.

The simplest example where coherence effects are important is a setup made of a single atom with spin $S=1/2$ precessing under an applied magnetic field in the presence of a spin-polarized tip.
If the polarizations of the applied field and of the tip are parallel, the rate equations yield the same results as our method.
In fact, in this case the process can essentially be described in a classical way.
However, our results show that if the tip polarization and the applied field are perpendicular, superposition effects are important and we find strong corrections to the rate equation results within the range of applicability of our approach.
Atoms with higher total spin, employed in the engineered nanomagnets, yield to qualitatively similar results that can be monitored by measuring the average current or the shot noise.
For more complex systems, such as spin chains, our results show that coherences contribute to the average current already at zero tip polarization. 

Although the present work only analyzes the steady state properties, coherence effects are crucial to describe the real time dynamics.
The present approach is therefore suitable to be applied to model the high-frequency magnetization dynamics observed in recent experiments \cite{Baumann2015,Krause2016}.
Calculation of the time dynamics will also allow us to make a comparison with numerically exact schemes such as the density matrix renormalization group \cite{Schollwock2005} and the quantum Monte Carlo method \cite{Antipov2016}.
It is also worthwhile to compare our results with the recently presented kinetic equation approach \cite{Maslova2016a,Maslova2016b}.
To summarize, the approach developed in this article provides a further step for the full quantum mechanical description of atomic spin devices and can therefore be used to explore new quantum coherent regimes that are of crucial importance if these systems are to be used for quantum information processing.   

\section*{Acknowledgments}
We gratefully acknowledge discussions with S. Otte, J. Fernandez-Rossier and A. Lichtenstein.
P.R. acknowledges support by FCT through Investigador FCT Contract No. IF/00347/2014. 
The method derivation (Sec. \ref{sec:Method}) was supported by RFBR Grant No. 16-32-00554.
The numerical modeling (Sec. \ref{sec:Results}) was funded by RSF Grant No. 16-42-01057.

\appendix

\section{Derivation of master equation}
\label{sec:Derivation}

In this appendix we explain some intermediate steps of the master equation derivation presented in Sec. \ref{sec:MethodDetails}.

\paragraph*{Derivation of Eq. (\ref{eqn:EOMinitial}).}

To obtain the equations of motion for CSDMs, we take the time derivative of Eq. (\ref{eqn:CSDMdef}) and use the von Neumann equation for the full density matrix
\begin{equation}
\partial_{t} {\rho}_{n}+i\mbox{tr}_{R}\left(\mathcal{P}_{n}\left[H,\rho_{\text{tot}}\right]\right)=0.
\label{eqn:App1}
\end{equation}
We substitute $ {H}= {H}_{A}+H_{R}+ {H}_{I}$ and use the commutativity of $\mathcal{P}_{n}$ with $ {H}_{A}$ and $H_{R}$ (these parts of the Hamiltonian do not generate the flow of particles between the leads) to show that
\begin{equation}
\mbox{tr}_{R}\left(\mathcal{P}_{n}\left[H_{A},\rho_{\text{tot}}\right]\right)=\left[H_{A}, {\rho}_{n}\right],~~~\mbox{tr}_{R}\left(\mathcal{P}_{n}\left[H_{R},\rho_{\text{tot}}\right]\right)=0.
\label{eqn:App2}
\end{equation}
Equation \ref{eqn:EOMinitial} then follows from Eq. \ref{eqn:App1}.

\paragraph*{Derivation of Eq. (\ref{eqn:EOMnonliouv}).}

The substitution of $ {H}_{I}=\sum_{\eta\eta'}\sqrt{J_{\eta}J_{\eta'}}\sum_{\alpha\alpha'} {T}_{\alpha\alpha'}c^{\dag}_{\alpha}c_{\alpha'}$ into the right-hand side of the equation of motion for CSDMs (\ref{eqn:EOMinitial}) gives
\begin{equation}
\begin{split}
&-i\sum_{\eta\eta'}\sqrt{J_{\eta}J_{\eta'}}\sum_{\alpha\alpha'}\mbox{tr}_{R}\left(\mathcal{P}_{n}\left[ {T}_{\alpha\alpha'}c^{\dag}_{\alpha}c_{\alpha'},\rho_{\text{tot}}\right]\right)=\\
&=\sum_{\eta\eta'}\sqrt{J_{\eta}J_{\eta'}}\sum_{\alpha\alpha'}\left(-i {T}_{\alpha\alpha'}\mbox{tr}_{R}\left(c^{\dag}_{\alpha}c_{\alpha'}\rho_{\text{tot}}\mathcal{P}_{n}\right)+\mbox{h.c.}\right).
\end{split}
\label{eqn:App3}
\end{equation}
We use the definition (\ref{eqn:C-operators}) to get
\begin{equation}
-i\mbox{tr}_{R}\left(c^{\dag}_{\alpha}c_{\alpha'}\rho_{\text{tot}}\mathcal{P}_{n}\right)= {C}^{(n)}_{\alpha\alpha'}-if_{\alpha}\delta_{\alpha\alpha'} {\rho}_{n}
\label{eqn:App4}
\end{equation}
and substitute this relation into Eq. (\ref{eqn:App3}).
After rearrangement of terms, one obtains Eq. (\ref{eqn:EOMnonliouv}).

\section{Operators $ {C}_{\alpha\alpha'}^{(n)}$}
\label{sec:C-operators}

This appendix contains the derivations of exact and approximate equations of motion (EOMs) for the auxiliary operators $ {C}^{(n)}_{\alpha\alpha'}$ and the solution of the approximate EOMs.

\paragraph*{Derivation of Eq. (\ref{eqn:ExactEOMforC}).}

To obtain the exact EOM for $ {C}_{\alpha\alpha'}^{(n)}$, we take the time derivative of the definition (\ref{eqn:C-operators}) and use the von Neumann equation for the full density matrix
\begin{equation}
\partial_{t} {C}_{\alpha\alpha'}^{(n)}=-\mbox{tr}_{R}\left(\left(c_{\alpha}^{\dag}c_{\alpha'}-f_{\alpha}\delta_{\alpha\alpha'}\right)\left[H,\rho_{\text{tot}}\right]\mathcal{P}_{n}\right).
\label{eqn:App5}
\end{equation}
With $ {H}= {H}_{A}+H_{R}+ {H}_{I}$ one gets 
\begin{equation}
\begin{split}
&\partial_{t} {C}_{\alpha\alpha'}^{(n)}+\mbox{tr}_{R}\left(\left(c_{\alpha}^{\dag}c_{\alpha'}-f_{\alpha}\delta_{\alpha\alpha'}\right)\left[ {H}_{A},\rho_{\text{tot}}\right]\mathcal{P}_{n}\right)+\\
&+\mbox{tr}_{R}\left(\left(c_{\alpha}^{\dag}c_{\alpha'}-f_{\alpha}\delta_{\alpha\alpha'}\right)\left[H_{R},\rho_{\text{tot}}\right]\mathcal{P}_{n}\right)=\\
&=-\mbox{tr}_{R}\left(\left(c_{\alpha}^{\dag}c_{\alpha'}-f_{\alpha}\delta_{\alpha\alpha'}\right)\left[H_{I},\rho_{\text{tot}}\right]\mathcal{P}_{n}\right).
\end{split}
\label{eqn:App6}
\end{equation}
For the second term on the left-hand side of this equation 
\begin{equation}
\begin{split}
&\mbox{tr}_{R}\left(\left(c_{\alpha}^{\dag}c_{\alpha'}-f_{\alpha}\delta_{\alpha\alpha'}\right)\left[H_{A},\rho_{\text{tot}}\right]\mathcal{P}_{n}\right)=\\
&=\left[H_{A},\mbox{tr}_{R}\left(\left(c_{\alpha}^{\dag}c_{\alpha'}-f_{\alpha}\delta_{\alpha\alpha'}\right)\rho_{\text{tot}}\mathcal{P}_{n}\right)\right]=i\left[ {H}_{A}, {C}_{\alpha\alpha'}^{(n)}\right].
\end{split}
\label{eqn:App7}
\end{equation}
For the third term we substitute $H_{R}$ and get 
\begin{equation}
\begin{split}
&\mbox{tr}_{R}\left(\left(c_{\alpha}^{\dag}c_{\alpha'}-f_{\alpha}\delta_{\alpha\alpha'}\right)\left[H_{R},\rho_{\text{tot}}\right]\mathcal{P}_{n}\right)=\\
&=\mbox{tr}_{R}\left(\left[c_{\alpha}^{\dag}c_{\alpha'},H_{R}\right]\rho_{\text{tot}}\mathcal{P}_{n}\right)=\\
&=\sum_{\beta}\varepsilon_{\beta}\mbox{tr}_{R}\left(\left[c_{\alpha}^{\dag}c_{\alpha'},c_{\beta}^{\dag}c_{\beta}\right]\rho_{\text{tot}}\mathcal{P}_{n}\right)=\\
&=-\left(\varepsilon_{\alpha}-\varepsilon_{\alpha'}\right)\mbox{tr}_{R}\left(c_{\alpha}^{\dag}c_{\alpha'}\rho_{\text{tot}}\mathcal{P}_{n}\right)=\\
&=-i\left(\varepsilon_{\alpha}-\varepsilon_{\alpha'}\right)C_{\alpha\alpha'}^{(n)},
\end{split}
\label{eqn:App8}
\end{equation}
where the relations $\left[c_{\alpha}^{\dag}c_{\alpha'},c_{\beta}^{\dag}c_{\beta}\right]=\delta_{\alpha'\beta}c_{\alpha}^{\dag}c_{\beta}-\delta_{\alpha\beta}c_{\beta}^{\dag}c_{\alpha'}$ and $\left(\varepsilon_{\alpha}-\varepsilon_{\alpha'}\right)\delta_{\alpha\alpha'}=0$ have been used.
The substitution of Eqs. (\ref{eqn:App7}) and (\ref{eqn:App8}) into Eq. (\ref{eqn:App6}) gives Eq. (\ref{eqn:ExactEOMforC}).

\paragraph*{Derivation of Eq. (\ref{eqn:ApproximateEOMforC}).}

Substituting $H_{I}$ into the right-hand side of Eq. (\ref{eqn:ExactEOMforC}) gives
\begin{equation}
\begin{split}
&-\mbox{tr}_{R}\left(\left(c_{\alpha}^{\dag}c_{\alpha'}-f_{\alpha}\delta_{\alpha\alpha'}\right)\left[H_{I},\rho_{\text{tot}}\right]\mathcal{P}_{n}\right)=\sum_{\mu\mu'}\sqrt{J_{\mu}J_{\mu'}}\times\\
&\times\sum_{\beta\beta'}\left(\mbox{tr}_{R}\left(\left(c_{\alpha}^{\dag}c_{\alpha'}-f_{\alpha}\delta_{\alpha\alpha'}\right)\rho_{\text{tot}}c_{\beta}^{\dag}c_{\beta'}\mathcal{P}_{n}\right) {T}_{\beta\beta'}-\right.\\
&\left.- {T}_{\beta\beta'}\mbox{tr}_{R}\left(\left(c_{\alpha}^{\dag}c_{\alpha'}-f_{\alpha}\delta_{\alpha\alpha'}\right)c_{\beta}^{\dag}c_{\beta'}\rho_{\text{tot}}\mathcal{P}_{n}\right)\right)=\\
&=\sum_{\mu\mu'}\sqrt{J_{\mu}J_{\mu'}}\sum_{\beta\beta'}\left(\mbox{tr}_{R}\left(c_{\beta}^{\dag}c_{\beta'}\left(c_{\alpha}^{\dag}c_{\alpha'}-f_{\alpha}\delta_{\alpha\alpha'}\right)\times\right.\right.\\
&\left.\times\mathcal{P}_{n-n_{\alpha\alpha'}}\rho_{\text{tot}}\mathcal{P}_{n+n_{\beta\beta'}}\right) {T}_{\beta\beta'}- {T}_{\beta\beta'}\mbox{tr}_{R}\left(\left(c_{\alpha}^{\dag}c_{\alpha'}-f_{\alpha}\delta_{\alpha\alpha'}\right)\times\right.\\
&\left.\left.\times c_{\beta}^{\dag}c_{\beta'}\mathcal{P}_{n-n_{\alpha\alpha'}-n_{\beta\beta'}}\rho_{\text{tot}}\mathcal{P}_{n}\right)\right),
\end{split}
\label{eqn:App9}
\end{equation}
where $n_{\alpha\alpha'}=\pm1,0$ is the number of electrons transferred from the tip to the substrate by $c_{\alpha}^{\dag}c_{\alpha'}$ operator.
We have used the identity $\mathcal{P}_{n}=\mathcal{P}_{n}^{2}$ and the commutation relation $\mathcal{P}_{n}c_{\alpha}^{\dag}c_{\alpha'}=c_{\alpha}^{\dag}c_{\alpha'}\mathcal{P}_{n-n_{\alpha\alpha'}}$.
As explained in the main text, we neglect components $\mathcal{P}_{m}\rho_{\text{tot}}\mathcal{P}_{n}$ of the full density matrix with $m\neq n$ and assume separability $\mathcal{P}_{n}\rho_{\text{tot}}\mathcal{P}_{n}\approx {\rho}_{n}\otimes\rho_{R}$ to approximate 
\begin{equation}
\begin{split}
&-\mbox{tr}_{R}\left(\left(c_{\alpha}^{\dag}c_{\alpha'}-f_{\alpha}\delta_{\alpha\alpha'}\right)\left[H_{I},\rho_{\text{tot}}\right]\mathcal{P}_{n}\right)\approx\sum_{\mu\mu'}\sqrt{J_{\mu}J_{\mu'}}\times\\
&\times\sum_{\beta\beta'}\delta_{n_{\alpha\alpha'},-n_{\beta\beta'}}\left(\left\langle c_{\beta}^{\dag}c_{\beta'}\left(c_{\alpha}^{\dag}c_{\alpha'}-f_{\alpha}\delta_{\alpha\alpha'}\right)\right\rangle\times {\rho}_{n-n_{\alpha\alpha'}} {T}_{\beta\beta'}-\right.\\
&\left.-\left\langle\left(c_{\alpha}^{\dag}c_{\alpha'}-f_{\alpha}\delta_{\alpha\alpha'}\right)c_{\beta}^{\dag}c_{\beta'}\right\rangle {T}_{\beta\beta'} {\rho}_{n}\right)=\sqrt{J_{\eta}J_{\eta'}}\times\\
&\times\left(\left(1-f_{\alpha}\right)f_{\alpha'} {\rho}_{n-n_{\alpha\alpha'}} {T}_{\alpha'\alpha}-f_{\alpha}\left(1-f_{\alpha'}\right) {T}_{\alpha'\alpha} {\rho}_{n}\right),
\end{split}
\label{eqn:App10}
\end{equation}
where $\left\langle\cdot\right\rangle=\mbox{tr}_{R}\left(\cdot\rho_{R}\right)$, and we have used the relations 
\begin{equation}
\begin{split}
&\left\langle\left(c_{\alpha}^{\dag}c_{\alpha'}-f_{\alpha}\delta_{\alpha\alpha'}\right)c_{\beta}^{\dag}c_{\beta'}\right\rangle=f_{\alpha}\left(1-f_{\alpha'}\right)\delta_{\alpha\beta'}\delta_{\alpha'\beta},\\
&\left\langle c_{\beta}^{\dag}c_{\beta'}\left(c_{\alpha}^{\dag}c_{\alpha'}-f_{\alpha}\delta_{\alpha\alpha'}\right)\right\rangle=\left(1-f_{\alpha}\right)f_{\alpha'}\delta_{\alpha\beta'}\delta_{\alpha'\beta}.
\end{split}
\label{eqn:App11}
\end{equation}
We note that $ {T}_{\alpha'\alpha}= {T}^{\dag}_{\alpha\alpha'}$ and obtain Eq. (\ref{eqn:ApproximateEOMforC}).

\paragraph*{Derivation of Eq. (\ref{eqn:ApproximateEOMforCSolution}).}

To solve Eq. (\ref{eqn:ApproximateEOMforC}), we use the substitution 
\begin{equation}
 {C}_{\alpha\alpha'}^{(n)}=e^{-i {H}_{A}t}\tilde{ {C}}_{\alpha\alpha'}^{(n)}e^{i {H}_{A}t}e^{i(\varepsilon_{\alpha}-\varepsilon_{\alpha'})t}.
\label{eqn:App12}
\end{equation}
One may show that $\tilde{ {C}}_{\alpha\alpha'}^{(n)}$ satisfies the equation 
\begin{equation}
\begin{split}
&\partial_{t}\tilde{ {C}}_{\alpha\alpha'}^{(n)}=\sqrt{J_{\eta}J_{\eta'}}e^{i {H}_{A}t}\left(\left(1-f_{\alpha}\right)f_{\alpha'} {\rho}_{n-n_{\alpha\alpha'}} {T}_{\alpha\alpha'}^{\dag}-\right.\\
&\left.-f_{\alpha}\left(1-f_{\alpha'}\right) {T}_{\alpha\alpha'}^{\dag} {\rho}_{n}\right)e^{-i {H}_{A}t}e^{-i(\varepsilon_{\alpha}-\varepsilon_{\alpha'})t}
\end{split}
\label{eqn:App13}
\end{equation}
with the initial condition 
\begin{equation}
\begin{split}
&\tilde{ {C}}_{\alpha\alpha'}^{(n)}(0)= {C}_{\alpha\alpha'}^{(n)}(0)=\\
&=-i\mbox{tr}_{R}\left(\left(c_{\alpha}^{\dag}c_{\alpha'}-f_{\alpha}\delta_{\alpha\alpha'}\right)\rho_{\text{tot}}(0)\mathcal{P}_{n}\right)=\\
&=-i\mbox{tr}_{R}\left(\left(c_{\alpha}^{\dag}c_{\alpha'}-f_{\alpha}\delta_{\alpha\alpha'}\right)\mathcal{P}_{n-n_{\alpha\alpha'}}\rho_{\text{tot}}(0)\mathcal{P}_{n}\right)=\\
&=-i\delta_{n,0}\delta_{n_{\alpha\alpha'},0} {\rho}(0)\left\langle c^{\dag}_{\alpha}c_{\alpha'}-f_{\alpha}\delta_{\alpha\alpha'}\right\rangle=0.
\end{split}
\label{eqn:App14}
\end{equation}
We have used the relation $\rho_{\text{tot}}(0)= {\rho}(0)\otimes\rho_{R}$ and the fact that no electrons are transferred at $t=0$.
The solution of Eq. (\ref{eqn:App13}) is then
\begin{equation}
\begin{split} & \tilde{ {C}}_{\alpha\alpha'}^{(n)}=\sqrt{J_{\eta}J_{\eta'}}\int\limits _{0}^{t}e^{i {H}_{A}t'}\left(\left(1-f_{\alpha}\right)f_{\alpha'} {\rho}_{n-n_{\alpha\alpha'}}(t') {T}_{\alpha\alpha'}^{\dag}-\right.\\
&\left.-f_{\alpha}\left(1-f_{\alpha'}\right) {T}_{\alpha\alpha'}^{\dag} {\rho}_{n}(t')\right)e^{-i {H}_{A}t'}e^{-i(\varepsilon_{\alpha}-\varepsilon_{\alpha'})t'}dt'.
\end{split}
\label{eqn:App15}
\end{equation}
Rotating back to $ {C}_{\alpha\alpha'}^{(n)}$ and introducing $\tau=t-t'$ gives Eq. (\ref{eqn:ApproximateEOMforCSolution}).

\section{Equation of motion (\ref{eqn:MEforCSDM})}
\label{sec:EOMforCSDM}

In this appendix we derive the equation of motion for charge-specific density matrices (\ref{eqn:MEforCSDM}).
It is obtained by substituting Eq. (\ref{eqn:Cfinal}) into Eq. (\ref{eqn:ApproximateEOMforC}), which results in
\begin{equation}
\begin{split}
&\partial_{t} {\rho}_{n}+i\left[ {H}_{A}+\sum_{\eta\alpha}J_{\eta}f_{\alpha} {T}_{\alpha\alpha}, {\rho}_{n}\right]=\sum_{\eta\eta'}J_{\eta}J_{\eta'}\times\\
&\times\sum_{\alpha\alpha'}\left(\left(1-f_{\alpha}\right)f_{\alpha'} {T}_{\alpha\alpha'} {\rho}_{n-n_{\alpha\alpha'}} {\mathcal{T}}^{\dag}_{\alpha\alpha'}-\right.\\
&\left.-f_{\alpha}(1-f_{\alpha'}) {T}_{\alpha\alpha'} {\mathcal{T}}^{\dag}_{\alpha\alpha'} {\rho}_{n}+\mbox{h.c.}\right).
\end{split}
\label{eqn:App16}
\end{equation}
The right-hand side of this equation depends on $ {\rho}_{n}$, $ {\rho}_{n-1}$, and $ {\rho}_{n+1}$.
We split it into two parts $ {F}_{1}+ {F}_{2}$, where the first one only depends on $ {\rho}_{n}$ as follows:
\begin{equation}
\begin{split}
& {F}_{1}=\sum_{\eta\eta'}J_{\eta}J_{\eta'}\sum_{\alpha\alpha'}\left(\left(1-f_{\alpha}\right)f_{\alpha'} {T}_{\alpha\alpha'} {\rho}_{n} {\mathcal{T}}^{\dag}_{\alpha\alpha'}-\right.\\
&\left.-f_{\alpha}(1-f_{\alpha'}) {T}_{\alpha\alpha'} {\mathcal{T}}^{\dag}_{\alpha\alpha'} {\rho}_{n}+\mbox{h.c.}\right),
\end{split}
\label{eqn:App17}
\end{equation}
and the second one is given by
\begin{equation}
\begin{split}
& {F}_{2}=\sum_{\eta\eta'}J_{\eta}J_{\eta'}\sum_{\alpha\alpha'}\left(1-f_{\alpha}\right)f_{\alpha'}\times\\
&\times\left( {T}_{\alpha\alpha'}\left( {\rho}_{n-n_{\alpha\alpha'}}- {\rho}_{n}\right) {\mathcal{T}}^{\dag}_{\alpha\alpha'}+\mbox{h.c.}\right).
\end{split}
\label{eqn:App18}
\end{equation}
We transform Eq. (\ref{eqn:App17}) as
\begin{equation}
\begin{split}
& {F}_{1}=\sum_{\eta\eta'}J_{\eta}J_{\eta'}\sum_{\alpha\alpha'}\left(1-f_{\alpha}\right)f_{\alpha'}\times\\
&\times\left( {\mathcal{T}}_{\alpha\alpha'} {\rho}_{n} {T}^{\dag}_{\alpha\alpha'}- {T}^{\dag}_{\alpha\alpha'} {\mathcal{T}}_{\alpha\alpha'} {\rho}_{n}+\mbox{h.c.}\right)=\\
&=\sum_{\eta\eta'}J_{\eta}J_{\eta'}\sum_{\alpha\alpha'}\left(1-f_{\alpha}\right)f_{\alpha'}\left( {\mathcal{T}}_{\alpha\alpha'} {\rho}_{n} {T}^{\dag}_{\alpha\alpha'}-\right.\\
&\left.-\frac{1}{2}\left\{ {T}^{\dag}_{\alpha\alpha'} {\mathcal{T}}_{\alpha\alpha'}, {\rho}_{n}\right\}+\mbox{h.c.}\right)-i\sum_{\eta\eta'}J_{\eta}J_{\eta'}\times\\
&\times\sum_{\alpha\alpha'}\left(1-f_{\alpha}\right)f_{\alpha'}\left[\frac{1}{2i}\left( {T}^{\dag}_{\alpha\alpha'} {\mathcal{T}}_{\alpha\alpha'}- {\mathcal{T}}^{\dag}_{\alpha\alpha'} {T}_{\alpha\alpha'}\right), {\rho}_{n}\right].
\end{split}
\label{eqn:App19}
\end{equation}
One may easily check that
\begin{equation}
 {F}_{1}-i\left[ {H}_{A}+\sum_{\eta\alpha}J_{\eta}f_{\alpha} {T}_{\alpha\alpha}, {\rho}_{n}\right]=\mathcal{L} {\rho}_{n}
\label{eqn:App20}
\end{equation}
with the superoperator $\mathcal{L}$ defined in Eq. (\ref{eqn:L-superoperatorDef}); thus the equation of motion (\ref{eqn:App16}) becomes
\begin{equation}
\partial_{t} {\rho}_{n}=\mathcal{L} {\rho}_{n}+ {F}_{2}.
\label{eqn:App21}
\end{equation}
The expression (\ref{eqn:App18}) only contains terms with $n_{\alpha\alpha'}\neq0$ when either $\eta=T$, $\eta'=S$ or $\eta=S$, $\eta'=T$.
We thus obtain
\begin{equation}
\begin{split}
& {F}_{2}=J_{T}J_{S}\sum_{st}\left(1-f_{s}\right)f_{t}\left( {T}_{st}\left( {\rho}_{n-1}- {\rho}_{n}\right) {\mathcal{T}}^{\dag}_{st}+\mbox{h.c.}\right)+\\
&+J_{T}J_{S}\sum_{st}\left(1-f_{t}\right)f_{s}\left( {T}_{ts}\left( {\rho}_{n+1}- {\rho}_{n}\right) {\mathcal{T}}^{\dag}_{ts}+\mbox{h.c.}\right),
\end{split}
\label{eqn:App22}
\end{equation}
where indices $t$ and $s$ enumerate the electronic states in the tip and the substrate correspondingly, and $n_{st}=1$, $n_{ts}=-1$ are used.
With definitions (\ref{eqn:DiscreteDerivatives}), (\ref{eqn:DJ-definition}), and (\ref{eqn:D-superoperators}) the last relation simplifies to
\begin{equation}
\begin{split}
& {F}_{2}=\mathcal{D}_{+}\left( {\rho}_{n-1}- {\rho}_{n}\right)+\mathcal{D}_{-}\left( {\rho}_{n+1}- {\rho}_{n}\right)=\\
&=-\mathcal{J} {\rho}'_{n}+\mathcal{D} {\rho}''_{n},
\end{split}
\label{eqn:App23}
\end{equation}
and Eq. (\ref{eqn:MEforCSDM}) is recovered from Eq. \ref{eqn:App21}.

\section{Wide band approximation}
\label{sec:WBA}

This appendix contains the derivation of Eq. (\ref{eqn:SumOverMomenta}).
We denote the required sum as $ {\Sigma}_{\lambda\lambda'}$ and calculate its matrix elements in the eigenbasis
\begin{equation}
\left\langle m\left| {\Sigma}_{\lambda\lambda'}\right|n\right\rangle=\sum_{kk'}\left(1-f_{\alpha}\right)f_{\alpha'}\left\langle m\left| {\mathcal{T}}_{\alpha\alpha'}\right|n\right\rangle.
\label{eqn:App24}
\end{equation}
Within the wide band approximation the sums over momenta translate to integrals according to the rule
\begin{equation}
\sum_{kk'}\rightarrow\varrho_{\lambda}\varrho_{\lambda'}\mathcal{V}_{\lambda}\mathcal{V}_{\lambda'}\iint\limits_{-W}^{W}d\varepsilon d\varepsilon'.
\label{eqn:App25}
\end{equation}
Substituting Eq. (\ref{eqn:NewOpsElements}) into Eq. (\ref{eqn:App24}), we thus obtain
\begin{equation}
\begin{split}
&\left\langle m\left| {\Sigma}_{\lambda\lambda'}\right|n\right\rangle=\varrho_{\lambda}\varrho_{\lambda'}\mathcal{V}_{\lambda}\mathcal{V}_{\lambda'}\left\langle m\left| {T}_{\lambda\lambda'}\right|n\right\rangle\left(\pi\iint\limits_{-W}^{W}d\varepsilon d\varepsilon'\times\right.\\
&\times\left(1-f_{\lambda}\left(\varepsilon\right)\right)f_{\lambda'}\left(\varepsilon'\right)\delta\left(\varepsilon-\varepsilon'+E_{m}-E_{n}\right)-\\
&\left.-iP\iint\limits_{-W}^{W}\frac{\left(1-f_{\lambda}\left(\varepsilon\right)\right)f_{\lambda'}\left(\varepsilon'\right)}{\varepsilon-\varepsilon'+E_{m}-E_{n}}d\varepsilon d\varepsilon'\right).
\end{split}
\end{equation}
Performing integration with equilibrium distribution functions $f_{\lambda}\left(\varepsilon\right)=\left(\exp\left(\beta\left(\varepsilon-\mu_{\lambda}\right)\right)+1\right)^{-1}$ in the large-$W$ limit results in
\begin{equation}
\begin{split}
&\left\langle m\left| {\Sigma}_{\lambda\lambda'}\right|n\right\rangle=\varrho_{\lambda}\varrho_{\lambda'}\mathcal{V}_{\lambda}\mathcal{V}_{\lambda'}\left\langle m\left|T_{\lambda\lambda'}\right|n\right\rangle\times\\
&\times\left(\frac{\pi}{\beta}g\left(\beta\left(\mu_{\lambda}-\mu_{\lambda'}+E_{m}-E_{n}\right)\right)-iW\ln4+\right.\\
&\left.+i\left(\mu_{\lambda}-\mu_{\lambda'}+E_{m}-E_{n}\right)\ln\frac{2\beta W}{\pi}\right).
\end{split}
\end{equation}
Finally, we use $\left(E_{m}-E_{n}\right)\left\langle m\left|T_{\lambda\lambda'}\right|n\right\rangle=\left\langle m\left|\left[H_{A},T_{\lambda\lambda'}\right]\right|n\right\rangle$ and recover $ {\Sigma}_{\lambda\lambda'}$ from its matrix elements to obtain Eq. (\ref{eqn:SumOverMomenta}).

\section{Inversion of $\mathcal{L}$}
\label{sec:Inversion}

This appendix explains the inversion procedure for the superoperator $\mathcal{L}$ that has to be performed to calculate the shot noise according to Eq. (\ref{eqn:summaryShotNoise}).
Let us consider a diagonalizable superoperator $\mathcal{L}$ with a unique stationary state.
We denote by $\lambda_{\alpha}$ the eigenvalues of $\mathcal{L}$ corresponding to the right and left eigenvectors $\chi_{\alpha}$ and $\tilde{\chi}_{\alpha}$ respectively, such that $\mathcal{L}\chi_{\alpha}=\lambda_{\alpha}\chi_{\alpha}$ and $\mathcal{L}^{\dagger}\tilde{\chi}_{\alpha}^{\dagger}=\bar{\lambda}_{\alpha}\tilde{\chi}_{\alpha}^{\dagger}$.
It is useful to use a bra-ket-like notation for which the preceding relations translate to 
\begin{equation}
\begin{split}
&\mathcal{L}\left|\chi_{\alpha}\right)=\lambda_{\alpha}\left|\chi_{\alpha}\right),\\
&\left(\tilde{\chi}_{\alpha}\right|\mathcal{L}=\left(\tilde{\chi}_{\alpha}\right|\lambda_{\alpha}.
\end{split}
\end{equation}
The eigenvectors can be chosen to respect the normalization condition
\begin{equation}
\left(\tilde{\chi}_{\alpha}|\chi_{\alpha'}\right)=\delta_{\alpha\alpha'},
\end{equation}
where the inner product is defined by 
\begin{equation}
\left(\tilde{\chi}_{\alpha}|\chi_{\alpha'}\right)=\sum_{mn}\bra n\tilde{\chi}_{\alpha}\ket m\bra m\chi_{\alpha'}\ket n.
\end{equation}
It follows from the fact that $\mathcal{L}$ is diagonalizable that its eigenvectors form a complete basis 
\begin{equation}
\left(a|b\right)=\sum_{\alpha}\left(a|\chi_{\alpha}\right)\left(\tilde{\chi}_{\alpha}|b\right)
\end{equation}
for generic matrices $a$ and $b$.
In this basis we also have 
\begin{equation}
f\left(\mathcal{L}\right)=\sum_{\alpha}\left|\chi_{\alpha}\right)f\left(\lambda_{\alpha}\right)\left(\tilde{\chi}_{\alpha}\right|.
\end{equation}
for an arbitrary function $f$.

The steady state $ {\rho}_{\infty}=\chi_{0}$ is the right eigenstate with zero eigenvalue.
On the other hand, since $\mathcal{L}$ is trace preserving, it has a left eigenvalue $\tilde{\chi}_{0}$ such that
$\bra n\tilde{\chi}_{0}\ket m=\delta_{nm}$.
Note that $\left(\tilde{\chi}_{0}|a\right)=\tr\left(a\right)$.
Using this notation, Eq. (\ref{MEforDM1_2}) can be written as 
\begin{equation}
\mathcal{L}\left| {\rho}_{\infty}^{(1)}\right)=\left(1-\left|\chi_{0}\right)\left(\tilde{\chi}_{0}\right|\right)\mathcal{J}\left|\chi_{0}\right)
\end{equation}
and thus, as the right-hand side has no component corresponding to the zero eigenspace of $\mathcal{L}$, the operator can be inverted as in Eq. (\ref{eqn:rho1steady}).

\section{Formulas for EASD}
\label{sec:EASD}

In this appendix we derive Eqs. (\ref{eqn:HamilShiftEASD}), (\ref{eqn:LsuperEASD}), (\ref{eqn:DsuperEASD}) that determine dynamics of EASDs, as explained in Sec. \ref{sec:MethodSummary}, and are used in the calculations in Sec. \ref{sec:Results}.
The structure of Eq. (\ref{eqn:EASDjumpopers}) implies that the substitution of operators $ {T}_{\lambda\lambda'}$ into Eqs. (\ref{eqn:HamilShiftSummed}), (\ref{eqn:LsuperSummed}), (\ref{eqn:DsuperSummed}) should be made according to the rules
\begin{equation}
\begin{split}
& T_{\lambda\lambda'}\rightarrow {S}_{r\sigma\sigma'},\\
& \sum_{\lambda\lambda'}\rightarrow\sum_{r\sigma\sigma'}\sum_{\eta\eta'}\left(\delta_{rr_{0}}+\left(1-\delta_{rr_{0}}\right)\delta_{\eta S}\delta_{\eta'S}\right).
\end{split}
\label{eqn:AppEASD1}
\end{equation}

\paragraph*{Derivation of Eq. (\ref{eqn:HamilShiftEASD}).}

Let us evaluate four parts of the Hamiltonian shift (\ref{eqn:HamilShiftSummed}) separately.
For the first part
\begin{equation}
\begin{split}
& \Delta_{1}H_{A}=\frac{W}{\pi}\sum_{\lambda}\gamma_{\lambda}T_{\lambda\lambda}=\frac{W}{\pi}\sum_{\eta r\sigma}\left(\delta_{rr_{0}}+\left(1-\delta_{rr_{0}}\right)\delta_{\eta S}\right)\times\\
& \times\gamma_{\eta\sigma}S_{r\sigma\sigma}=\frac{W}{\pi}\left(\sum_{\sigma}\gamma_{T\sigma}S_{r_{0}\sigma\sigma}+\sum_{r\sigma}\gamma_{S\sigma}S_{r\sigma\sigma}\right),
\end{split}
\label{eqn:AppEASD2}
\end{equation}
where we have summed over $\eta=T,S$.
We note that $\gamma_{S\sigma}=\gamma_{S}$ and $\gamma_{T\sigma}=w_{\sigma}\gamma_{T}$ due to the spin-dependent density of states.
Using Eq. (\ref{eqn:Soperators}), we obtain
\begin{equation}
\Delta_{1}H_{A}=\frac{W}{\pi}\gamma_{T}\left(w_{\uparrow}-w_{\downarrow}\right)S_{r_{0}z}=\frac{W}{\pi}p\gamma_{T}S_{r_{0}z}.
\label{eqn:AppEASD3}
\end{equation}
To calculate the second part of $\Delta H_{A}$, we present it in the form
\begin{equation}
\begin{split}
& \Delta_{2} {H}_{A}=\frac{1}{\pi\beta}\frac{1}{2i}\left( {G}- {G}^{\dag}\right),\\
& G=\sum_{\lambda\lambda'}\gamma_{\lambda}\gamma_{\lambda'} {T}^{\dag}_{\lambda\lambda'} {T}'_{\lambda\lambda'}.
\end{split}
\label{eqn:AppEASD4}
\end{equation}
With the summation rules (\ref{eqn:AppEASD1}) we get
\begin{equation}
\begin{split}
& G=\sum_{r\sigma\sigma'}S^{\dag}_{r\sigma\sigma'}A_{r\sigma\sigma'},\\
& A_{r\sigma\sigma'}=\sum_{\eta\eta'}\left(\delta_{rr_{0}}+\left(1-\delta_{rr_{0}}\right)\delta_{\eta S}\delta_{\eta'S}\right)\gamma_{\eta\sigma}\gamma_{\eta'\sigma'}Q^{\eta\eta'}_{r\sigma\sigma'}.
\end{split}
\label{eqn:AppEASD5}
\end{equation}
where the auxiliary operators $Q^{\eta\eta'}_{r\sigma\sigma'}$ are defined through their matrix elements as
\begin{equation}
\begin{split}
& \left\langle m\left|Q^{\eta\eta'}_{r\sigma\sigma'}\right|n\right\rangle=\\
& =g\left(\beta\left(\mu_{\eta}-\mu_{\eta'}+E_{m}-E_{n}\right)\right)\left\langle m\left|S_{r\sigma\sigma'}\right|n\right\rangle.
\end{split}
\label{eqn:AppEASD6}
\end{equation}
When $r$, $\sigma$, and $\sigma'$ are fixed, this expression gives one of the three operators defined in Eq. (\ref{eqn:Qoperators}): (i) $Q^{(0)}_{r\sigma\sigma'}$ for $\eta=\eta'$, (ii) $Q^{(+)}_{r\sigma\sigma'}$ for $\eta=S$ and $\eta'=T$, (iii) $Q^{(-)}_{r\sigma\sigma'}$ for $\eta=T$ and $\eta'=S$.
Summation over $\eta,\eta'$ in Eq. (\ref{eqn:AppEASD5}) gives
\begin{equation}
\begin{split}
& A_{r\sigma\sigma'}=\left(\gamma_{S\sigma}\gamma_{S\sigma'}+\delta_{rr_{0}}\gamma_{T\sigma}\gamma_{T\sigma'}\right)Q^{(0)}_{r\sigma\sigma'}+\\
& +\delta_{rr_{0}}\left(\gamma_{S\sigma}\gamma_{T\sigma'}Q^{(+)}_{r\sigma\sigma'}+\gamma_{T\sigma}\gamma_{S\sigma'}Q^{(-)}_{r\sigma\sigma'}\right),
\end{split}
\label{eqn:AppEASD7}
\end{equation}
from which one may recover Eq. (\ref{eqn:Aoperators}).
For the third and fourth parts of $\Delta H_{A}$ we get
\begin{equation}
\begin{split}
& \Delta_{3}H_{A}=-\frac{W\ln4}{\pi^{2}}K, \\
& \Delta_{4}H_{A}=\frac{1}{2\pi^{2}}\ln\frac{2\beta W}{\pi}\left[H_{A},K\right], \\
& K=\sum_{\lambda\lambda'}\gamma_{\lambda}\gamma_{\lambda'}T^{\dag}_{\lambda\lambda'}T_{\lambda\lambda'}.
\end{split}
\label{eqn:AppEASD8}
\end{equation}
Evaluating the operator $K$ gives
\begin{equation}
\begin{split}
& K=\sum_{r\sigma\sigma'}S^{\dag}_{r\sigma\sigma'}S_{r\sigma\sigma'}\sum_{\eta\eta'}\gamma_{\eta\sigma}\gamma_{\eta'\sigma'}\left(\delta_{rr_{0}}+\left(1-\delta_{rr_{0}}\right)\times\right. \\
& \left.\times\delta_{\eta S}\delta_{\eta'S}\right)=\gamma^{2}_{S}\sum_{r\sigma\sigma'}S^{\dag}_{r\sigma\sigma'}S_{r\sigma\sigma'}+\gamma^{2}_{T}\sum_{\sigma\sigma'}w_{\sigma}w_{\sigma'}\times \\
& \times S^{\dag}_{r_{0}\sigma\sigma'}S_{r_{0}\sigma\sigma'}+\gamma_{S}\gamma_{T}\sum_{\sigma\sigma'}\left(w_{\sigma}+w_{\sigma'}\right)S^{\dag}_{r_{0}\sigma\sigma'}S_{r_{0}\sigma\sigma'}= \\
& =2\left(\gamma^{2}_{S}\sum_{r}\mathbf{S}^{2}_{r}+2\gamma_{S}\gamma_{T}\mathbf{S}^{2}_{r_{0}}+\right. \\
& \left.+\gamma^{2}_{T}\left(\left(1-p^{2}\right)\mathbf{S}^{2}_{r_{0}}+2p^{2}S^{2}_{r_{0}z}\right)\right),
\end{split}
\label{eqn:AppEASD9}
\end{equation}
where we used the relation $\sum_{\sigma\sigma'}S^{\dag}_{r\sigma\sigma'}S_{r\sigma\sigma'}=2\mathbf{S}^{2}_{r}$.
Summing Eqs. (\ref{eqn:AppEASD3}), (\ref{eqn:AppEASD4}), and (\ref{eqn:AppEASD8}) and using $\left[H_{A},\mathbf{S}^{2}_{r}\right]=0$, one may recover the Hamiltonian shift (\ref{eqn:HamilShiftEASD}).

\paragraph*{Derivation of Eq. (\ref{eqn:LsuperEASD}).}

Let us evaluate the non-Liouvillian part of Eq. (\ref{eqn:LsuperEASD}).
Similarly to the derivation of the expression for $ {G}$ in Eq. (\ref{eqn:AppEASD5}), one may show that
\begin{equation}
\sum_{\lambda\lambda'}\gamma_{\lambda}\gamma_{\lambda'}T'_{\lambda\lambda'}\chi T^{\dag}_{\lambda\lambda'}=\sum_{r\sigma\sigma'}A_{r\sigma\sigma'}\chi S^{\dag}_{r\sigma\sigma'}.
\label{eqn:AppEASD10}
\end{equation}
That leads us to Eq. (\ref{eqn:LsuperEASD}).

\paragraph*{Derivation of Eq. (\ref{eqn:DsuperEASD}).}

Finally, we evaluate the expressions (\ref{eqn:DsuperSummed}).
Since the tunneling between the tip and the substrate only happens through the atom at $ {r}_{0}$, we have the substitution rules $\sum_{\lambda_{S}\lambda_{T}}\rightarrow\sum_{\sigma\sigma'}$ and $ {T}_{\lambda_{S}\lambda_{T}}, {T}_{\lambda_{T}\lambda_{S}}\rightarrow {S}_{r_{0}\sigma\sigma'}$.
Moreover, we obtain $ {T}'_{\lambda_{S}\lambda_{T}}= {Q}^{(+)}_{r_{0}\sigma\sigma'}$ and $ {T}'_{\lambda_{T}\lambda_{S}}= {Q}^{(-)}_{r_{0}\sigma\sigma'}$.
Thus
\begin{equation}
\begin{split}
& \mathcal{D}_{+} {\chi}=\frac{1}{\pi\beta}\sum_{\sigma\sigma'}\gamma_{S\sigma}\gamma_{T\sigma'}\left( {Q}^{(+)}_{r_{0}\sigma\sigma'} {\chi} {S}^{\dag}_{r_{0}\sigma\sigma'}+\mbox{h.c.}\right),\\
& \mathcal{D}_{-} {\chi}=\frac{1}{\pi\beta}\sum_{\sigma\sigma'}\gamma_{T\sigma}\gamma_{S\sigma'}\left( {Q}^{(-)}_{r_{0}\sigma\sigma'} {\chi} {S}^{\dag}_{r_{0}\sigma\sigma'}+\mbox{h.c.}\right),
\end{split}
\label{eqn:AppEASD11}
\end{equation}
which is equivalent to Eq. (\ref{eqn:DsuperEASD}).

\section{Lindblad analysis}
\label{sec:Lindblad}

In this appendix we identify several cases when the superoperator (\ref{eqn:LsuperEASD}) of the master equation (\ref{eqn:MEforDM}) has the Lindblad form.
For this we compare the operators $ {A}_{r\sigma\sigma'}$ and $ {S}_{r\sigma\sigma'}$ whose matrix elements are related to each other by
\begin{equation}
\begin{split}
&\left\langle m\left| {A}_{r\sigma\sigma'}\right|n\right\rangle=g^{mn}_{r\sigma\sigma'}\left\langle m\left| {S}_{r\sigma\sigma'}\right|n\right\rangle,\\
&g^{mn}_{r\sigma\sigma'}=\delta_{rr_{0}}\gamma_{S}\gamma_{T}\left(w_{\sigma}g\left(\beta\left(E_{m}-E_{n}+eV\right)\right)+\right.\\
&\left.+w_{\sigma'}g\left(\beta\left(E_{m}-E_{n}-eV\right)\right)\right)+\left(\gamma^{2}_{S}+\delta_{rr_{0}}\gamma^{2}_{T}w_{\sigma}w_{\sigma'}\right)\times\\
&\times g\left(\beta\left(E_{m}-E_{n}\right)\right).
\end{split}
\label{eqn:AppLind1}
\end{equation}
The required proportionality relation $A_{r\sigma\sigma'}\sim S_{r\sigma\sigma'}$ is fulfilled when factors $g^{mn}_{r\sigma\sigma'}$ do not depend on the states $|m\rangle$ and $|n\rangle$ for all nonvanishing matrix elements $\left\langle m\left|S_{r\sigma\sigma'}\right|n\right\rangle$.
Below we consider the situations when this happens.

\paragraph*{Infinite temperature.}

In this case $\beta\rightarrow0$ and $g(x)\to1$ for all arguments of the function that occur in Eq. (\ref{eqn:AppLind1}), so that g-factors do not depend on $m$ and $n$.
We thus get $A_{r\sigma\sigma'}=a_{r\sigma\sigma'} {S}_{r\sigma\sigma'}$ with
\begin{equation}
a_{r\sigma\sigma'}=\gamma^{2}_{S}+\delta_{rr_{0}}\left(\gamma_{S}\gamma_{T}\left(w_{\sigma}+w_{\sigma'}\right)+\gamma^{2}_{T}w_{\sigma}w_{\sigma'}\right).
\label{eqn:AppLind2}
\end{equation}
The superoperator (\ref{eqn:LsuperEASD}) simplifies to
\begin{equation}
\begin{split}
& \mathcal{L} {\chi}=-i\left[H'_{A},\chi\right]+\frac{2}{\pi\beta}\sum_{r\sigma\sigma'}a_{r\sigma\sigma'}\times\\
& \times\left(S_{r\sigma\sigma'}\chi S^{\dag}_{r\sigma\sigma'}-\frac{1}{2}\left\{S^{\dag}_{r\sigma\sigma'}S_{r\sigma\sigma'},\chi\right\}\right).
\end{split}
\label{eqn:AppLind3}
\end{equation}
It has the Lindblad form with positive coefficients (\ref{eqn:AppLind2}).

\paragraph*{Infinite voltage.}

In this case some $g$ factors become much larger than others, and we only take them into account.
For large positive voltage $V>0$ we approximate
\begin{equation}
g(\beta(E_{m}-E_{n}-xeV))\approx
\begin{cases}
~0,~x=-1,0,\\
~\beta eV,~x=+1,
\end{cases}
\label{eqn:AppLind4}
\end{equation}
which leads to $Q^{(+)}_{r\sigma\sigma'}=Q^{(0)}_{r\sigma\sigma'}=0$ and $Q^{(-)}_{r\sigma\sigma'}=\beta eV S_{r\sigma\sigma'}$.
One thus gets $A_{r\sigma\sigma'}=\delta_{rr_{0}}\gamma_{S}\gamma_{T}w_{\sigma'}\beta eV S_{r\sigma\sigma'}$ and
\begin{equation}
\begin{split}
& \mathcal{L}\chi=-i\left[H'_{A},\chi\right]+\frac{2}{\pi}eV\gamma_{S}\gamma_{T}\sum_{\sigma\sigma'}w_{\sigma'}\times\\
& \times\left(S_{r_{0}\sigma\sigma'}\chi S^{\dag}_{r_{0}\sigma\sigma'}-\frac{1}{2}\left\{S^{\dag}_{r_{0}\sigma\sigma'}S_{r_{0}\sigma\sigma'},\chi\right\}\right).
\end{split}
\label{eqn:AppLind5}
\end{equation}
This superoperator has the Lindblad form with positive coefficients.
Analogously, for large negative voltage $V<0$
\begin{equation}
\begin{split}
& \mathcal{L}\chi=-i\left[H'_{A},\chi\right]+\frac{2}{\pi}e\left|V\right|\gamma_{S}\gamma_{T}\sum_{\sigma\sigma'}w_{\sigma}\times\\
& \times\left(S_{r_{0}\sigma\sigma'}\chi S^{\dag}_{r_{0}\sigma\sigma'}-\frac{1}{2}\left\{ S^{\dag}_{r_{0}\sigma\sigma'} S_{r_{0}\sigma\sigma'},\chi\right\}\right).
\end{split}
\end{equation}

\paragraph*{Single atom in parallel magnetic field.}

We consider the situation when there is no crystal anisotropy, and the spectrum of the atom is equidistant.
In the case of the parallel external magnetic field $\mathbf{B}\parallel\mathbf{P}$, all nonvanishing matrix elements of any operator $S_{\sigma\sigma'}$ have the same $E_{m}-E_{n}$.
In particular, (i) $\pm S_{z}$ requires $m=n$, so $E_{m}-E_{n}=0$, (ii) $ {S}^{+}$ requires $m=n+1$, so $E_{m}-E_{n}=\Delta$, (iii) $ {S}^{-}$ requires $m=n-1$, so $E_{m}-E_{n}=-\Delta$.
This leads to the proportionality relation $A_{\sigma\sigma'}=a_{\sigma\sigma'}S_{\sigma\sigma'}$ with positive coefficients
\begin{equation}
\begin{split}
&a_{\sigma\sigma}=\gamma_{S}\gamma_{T}w_{\sigma}\left(g\left(\beta eV\right)+g\left(-\beta eV\right)\right)+\\
&+\gamma^{2}_{S}+\gamma^{2}_{T}w^{2}_{\sigma},~\sigma=\uparrow,\downarrow,\\
&a_{\uparrow\downarrow}=\left(\gamma^{2}_{S}+\gamma^{2}_{T}w_{\uparrow}w_{\downarrow}\right)g\left(-\beta\Delta\right)+\\
&+\gamma_{S}\gamma_{T}\left(w_{\uparrow}g\left(\beta\left(-\Delta+eV\right)\right)+w_{\downarrow}g\left(\beta\left(-\Delta-eV\right)\right)\right),\\
&a_{\downarrow\uparrow}=\left(\gamma^{2}_{S}+\gamma^{2}_{T}w_{\downarrow}w_{\uparrow}\right)g\left(\beta\Delta\right)+\\
&+\gamma_{S}\gamma_{T}\left(w_{\downarrow}g\left(\beta\left(\Delta+eV\right)\right)+w_{\uparrow}g\left(\beta\left(\Delta-eV\right)\right)\right),
\end{split}
\end{equation}
so that the Lindblad form of the superoperator is recovered:
\begin{equation}
\begin{split}
& \mathcal{L} {\chi}=-i\left[H'_{A}, {\chi}\right]+\frac{2}{\pi\beta}\sum_{\sigma\sigma'}a_{\sigma\sigma'}\times\\
& \times\left(S_{\sigma\sigma'}\chi S^{\dag}_{\sigma\sigma'}-\frac{1}{2}\left\{S^{\dag}_{\sigma\sigma'}S_{\sigma\sigma'}, \chi\right\}\right).
\end{split}
\end{equation}

\section{Parallel geometry}
\label{sec:Parallel}

This appendix presents the results obtained for the steady state characteristics of single atoms with spin $S=1/2$ and $S=5/2$ in the case when the applied magnetic field $\mathbf{B}$ is parallel to the tip polarization $\mathbf{P}$ (both vectors are along $z$ axis).
The plots presented below are the same for both the ME and REs methods.

\begin{figure}[t]
\includegraphics[width=1\columnwidth]{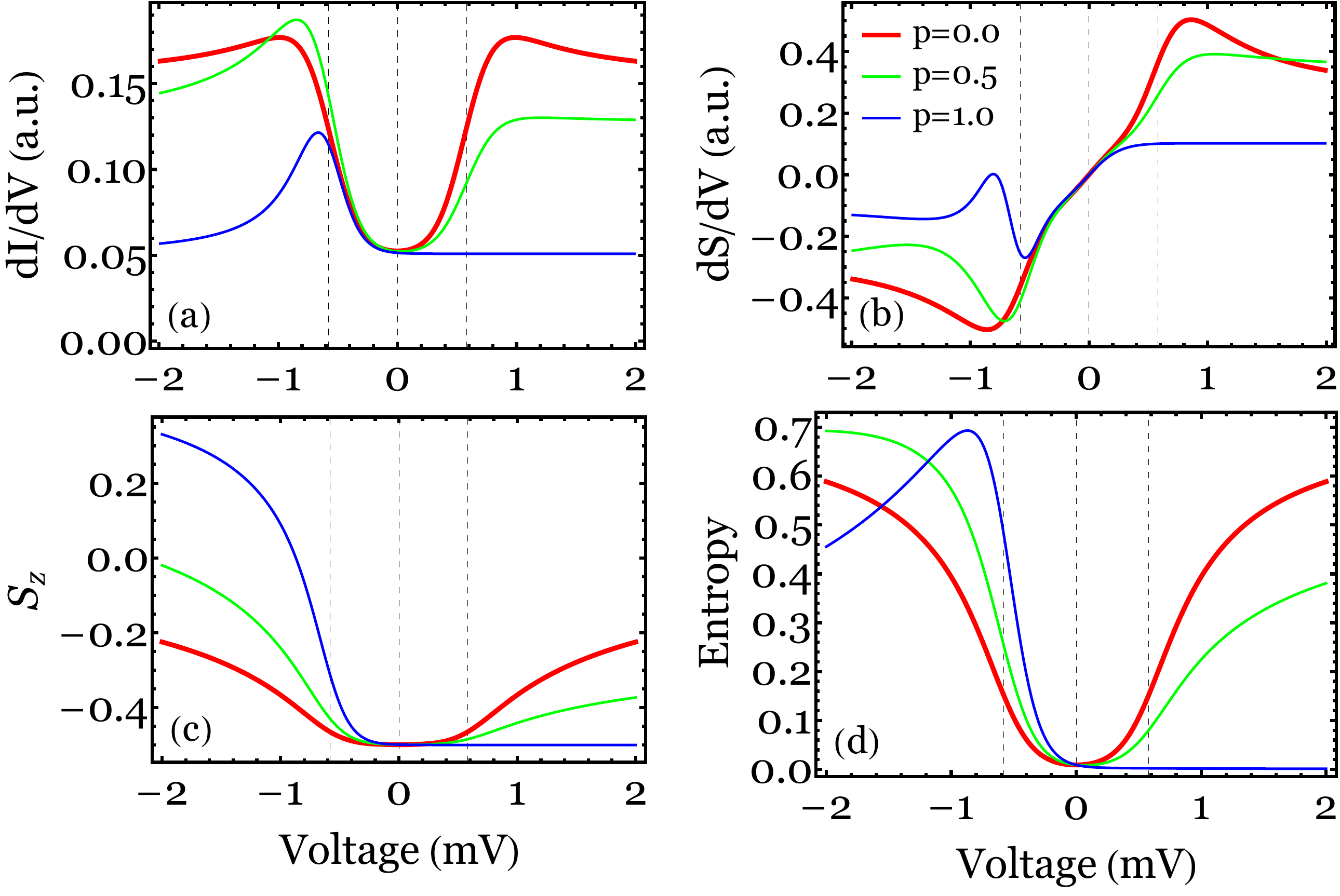}
\caption{Steady state characteristics for a single spin $S=1/2$ in a parallel geometry (both $\mathbf{B}$ and $\mathbf{P}$ along $z$ axis) for different values of the polarization parameter $p$ and for $g=2$, $B=5$ T, $\gamma_{T}=\gamma_{S}=0.8$, $(\beta k_{B})^{-1}=1$ K.
The quantities presented as functions of voltage are (a) differential conductance, (b) differential shot noise, (c) average spin component $\langle S_{z}\rangle$, (d) entropy.
In this geometry ME and RE approaches are equivalent, as the coherences vanish due to $\langle S_{x}\rangle=\langle S_{y}\rangle=0$.}
\label{fgr:SingleOneHalfParallel} 
\end{figure}

\paragraph*{Single spin $S=1/2$.}

The spectra of the steady state observables for a single spin $S=1/2$ are shown in Fig. \ref{fgr:SingleOneHalfParallel}.
Jumps in the differential conductance arise at $eV=\pm g\mu_{B}\abs{\mathbf{B}}$, as above this energy the inelastic conducting channel, which involves the spin-flip process, is energetically accessible.
As the polarization increases, the tip density of states becomes more and more spin asymmetric and processes that consist of tunneling a minority spin from or to the tip are suppressed.
In the limit of a fully polarized tip with $p=1$, the inelastic spin-flip channel only arises between majority spins in the tip, which explains the asymmetry of the $dI/dV$ curve.
We note that for substantially large values of voltage $|eV|>W$ the current must saturate and the differential conductance must approach zero.
However, this saturation effect cannot be seen in the presented results, since we consider the voltage range $|eV|\ll W$ in the derivations above.
The entanglement von Neumann entropy generically varies with the voltage but also depends on the tip polarization.
For unpolarized tip it increases with the voltage amplitude, while a fully polarized tip decreases the entropy by driving the atom into a pure spin-polarized state.

\begin{figure}[t]
\includegraphics[width=1\columnwidth]{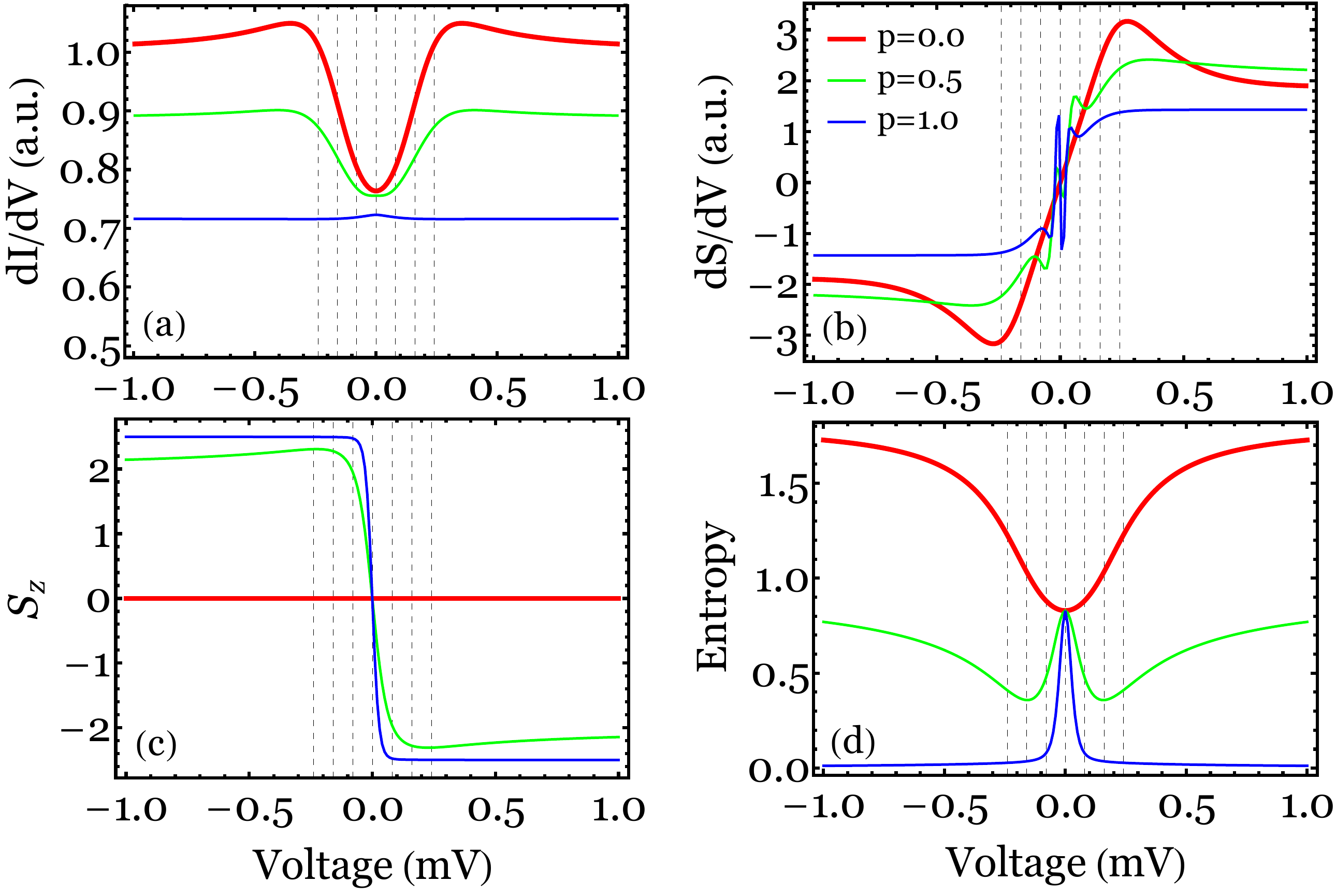}
\caption{Steady state characterization for a single spin $S=5/2$ in a parallel geometry ($\mathbf{P}$ along the easy axis $z$ of the crystal) for different values of the polarization parameter $p$ and for $D=-0.04$ meV, $\gamma_{T}=\gamma_{S}=0.6$, $(\beta k_{B})^{-1}=0.5$ K.
The quantities presented as functions of voltage are (a) differential conductance, (b) differential shot noise, (c) average spin component $\langle S_{z}\rangle$, (d) entropy.
Other components of the spin vanish, i.e., $\langle S_{x}\rangle=\langle S_{y}\rangle=0$.
As in the case of spin $S=1/2$, ME and RE approaches are equivalent in this geometry.}
\label{fgr:SingleFiveHalfParallel} 
\end{figure}

\paragraph*{Single spin $S=5/2$.}

The spectra of the steady state observables for a single spin $S=5/2$ are presented in Fig. \ref{fgr:SingleFiveHalfParallel}.
In this case the anisotropy is set by the crystal field yielding the energy levels of the atom to lie on a down-turned parabola.
As a result, one can see the characteristic switching between two degenerate ground states $S_{z}=5/2$ and $S_{z}=-5/2$.
This switching occurs via transitions to high-energy magnetic states with $S_{z}$ between these extreme values.
The required excitation energy is provided by the tunneling electrons that drive the atom to either the $S_{z}=5/2$ or $S_{z}=-5/2$ state depending on the polarity of the current.
This switching is observed in the voltage dependence of the average spin projection $\langle S_{z}\rangle$ and the entropy.
For the unpolarized tip with $p=0$, there is no switching and the entropy monotonically increases with the voltage.
For the polarized tip with $p\neq0$, the region of the voltage where the switching occurs is characterized by the entropy decrease.

\newpage
\bibliography{EASDBib}

\end{document}